\documentclass[manuscript,screen,nonacm]{acmart}

\usepackage{tcolorbox}
\usepackage{cleveref}
\usepackage{listings}
\usepackage{xspace}
\usepackage{colortbl}

\tcbset{colback=gray!5!white,colframe=gray!75!black,boxsep=3pt,left=3pt,right=3pt,top=3pt,bottom=3pt}

\definecolor{CustomYellow}{RGB}{255, 171, 64}
\newcommand{\nbRules}{30}
\newcommand{\circled}[2][CustomYellow]{\tikz[baseline=(char.base)]{\node[shape=circle, fill=#1, text=white, inner sep=1pt] (char) {\small #2};}}

\newcommand{\evo}{{\sc EvoMaster}\xspace}
\newcommand{\statictool}{\textsc{SCOAS}}

\definecolor{codegreen}{rgb}{0.25,0.5,0.35}
\definecolor{codegray}{rgb}{0.5,0.5,0.5}
\definecolor{codepurple}{rgb}{0.6,0,0}
\definecolor{backcolour}{rgb}{0.95,0.95,0.92}
\definecolor{colorstring}{rgb}{0.5,0,0.35}
\definecolor{rltred}{rgb}{0.5,0,0}
\definecolor{rltgreen}{rgb}{0,0.5,0}
\definecolor{rltblue}{rgb}{0,0,0.5}
\definecolor{DarkGreen}{rgb}{0.00,0.60,0.00}
\definecolor{ScarletRed}{rgb}{0.80,0.00,0.00}
\definecolor{blizzardblue}{rgb}{0.67, 0.9, 0.93}
\definecolor{green-yellow}{rgb}{0.68, 1.0, 0.18}
\definecolor{dkgreen}{rgb}{0,0.6,0}
\definecolor{gray}{rgb}{0.5,0.5,0.5}
\definecolor{mauve}{rgb}{0.58,0,0.82}
\definecolor{lightgrey}{rgb}{0.90,0.90,0.90}
\definecolor{grey}{gray}{0.75}
\definecolor{light-gray}{gray}{0.80}

\lstdefinestyle{mystyle}{
    escapechar=©, 
	backgroundcolor=\color{backcolour},
    basicstyle=\footnotesize\ttfamily,
   	identifierstyle=\footnotesize\ttfamily,
	commentstyle=\color{codegreen},
	keywordstyle=\color{colorstring}\bfseries,
	numberstyle=\ttfamily\color{codegray},
	stringstyle=\ttfamily\color{DarkGreen},
	breakatwhitespace=false,
	breaklines=true,
	captionpos=b,
	keepspaces=true,
	numbers=left, 
	numbersep=2pt,
	showspaces=false,
	showstringspaces=false,
	showtabs=false,
	tabsize=2
}
\begin{document}

\title{Detecting HTTP Status Code Misuses in REST APIs via Static and Dynamic Analysis}

\author{Alix Decrop}
\email{alix.decrop@unamur.be}
\orcid{0009-0007-2641-5983}
\affiliation{%
  \institution{NADI, University of Namur}
  \city{Namur}
  \country{Belgium}
}

\author{Andrea Arcuri}
\email{andrea.arcuri@kristiania.no}
\orcid{0000-0003-0799-2930}
\affiliation{%
  \institution{Kristiania University of Applied Sciences and Oslo Metropolitan University}
  \city{Oslo}
  \country{Norway}
}

\author{Mike Papadakis}
\email{michail.papadakis@uni.lu}
\orcid{0000-0003-1852-2547}
\affiliation{%
  \institution{SnT, University of Luxembourg}
  \city{Luxembourg}
  \country{Luxembourg}
}

\author{Pierre-Yves Schobbens}
\email{pierre-yves.schobbens@unamur.be}
\orcid{0000-0001-8677-4485}
\affiliation{%
  \institution{NADI, University of Namur}
  \city{Namur}
  \country{Belgium}
}

\author{Gilles Perrouin}
\email{gilles.perrouin@unamur.be}
\orcid{0000-0002-8431-0377}
\affiliation{%
  \institution{NADI, University of Namur}
  \city{Namur}
  \country{Belgium}
}

\renewcommand{\shortauthors}{Decrop et al.}

\begin{abstract}

REST APIs are widely used on the web for client-server communications. As REST is based on HTTP, server responses contain status codes to indicate the outcome of requests (e.g., \texttt{200 OK} for a success and \texttt{404 Not Found} for an unavailable resource). While HTTP status codes are standardized, their semantics are not enforced in REST, leading to many misuses in practice (e.g., using \texttt{500 Internal Server Error} to describe a client error). Such misuses may have nefarious consequences, such as reducing interoperability, misleading API clients, or causing false positives in testing tools. In this paper, we present a combined static and dynamic analysis approach for detecting HTTP status code misuses in REST APIs. We first study 2,625 real-world REST API specifications to identify relevant status codes and derive a set of 30 usage rules based on HTTP standards and REST API principles. We then implement tools to identify such rule violations in OpenAPI specifications (static analysis) and in API behavior (dynamic analysis). Our evaluation finds that status code misuses are frequent and systematic in REST APIs, with both static and dynamic approaches detecting various misuses. We highlight that both approaches may be used in a complementary manner, and also provide insight for REST API testers and users alike.

\end{abstract}



\maketitle

\section{Introduction}
\label{sec:introduction}

Many modern web Application Programming Interfaces (APIs) rely on the REpresentational State Transfer (REST) architectural style \cite{fielding2000architectural}. Such REST (also termed \textit{RESTful}) APIs handle client-server communications through HTTP. Via its comprehensive list of methods and status codes, the HTTP standard allows for interoperable communication among web services. However, this paper finds that many REST APIs deviate from HTTP semantics regarding the correct use of status codes; We call such deviations \textit{status codes misuses}.

Status code misuses can have several harmful consequences. For instance, some APIs returns status codes in the \texttt{5xx} range for client errors. However, when a client error is encountered, a code in the \texttt{4xx} range is expected, as \texttt{5xx} codes are exclusively reserved for server errors. This confuses clients, falsely believing that errors come from the server and not from their side. Similarly, misuses also mislead REST API analysis and testing tools, threatening the conclusions one can make from their usage and causing result inaccuracies (e.g., \texttt{5xx} codes are often directly treated as bugs by fuzzing tools). Additionally, APIs may implement their own status codes that do not map to official HTTP status codes interpretations, forcing clients to implement non-interoperable parsing if they wish to be more informative to their users. Furthermore, APIs may return overly generic status codes, such as \texttt{400 Bad Request} for an unsupported content format instead of \texttt{415 Unsupported Media Type}. This is detrimental to API users, as debugging invalid requests become a tedious task due to a lack of server information.

This paper investigates HTTP status code misuses in REST APIs, at a static and dynamic level. In particular, this paper makes the following novel contributions:

\begin{itemize}
    \item An evaluation dataset combining three state-of-the-art REST API datasets/benchmarks~\cite{apisguru2026apis,decrop2025public,sahin2025wfc}, comprising a total of 2,625 distinct API specifications and 36 API implementations.
    
    \item A study on the distribution of status codes in the REST APIs of the dataset.
    
    \item A set of \nbRules{} status code usage rules, derived from official HTTP semantics and REST API design/principles.
    
    \item \statictool{} (Status Code Analysis in OpenAPI Specifications), a static analysis tool aimed at detecting status code misuses in OpenAPI specifications.

    \item An extension of the tool \evo~\cite{arcuri2019restful,arcuri2025tool}, providing a dynamic analysis of status code misuses in REST API implementations.

    \item An evaluation of status code misuse detection with \statictool{} and \evo.

    \item Our implementation and evaluation data, which can be found in the data availability statement at the end of the work.
\end{itemize}

The tool \statictool{} was introduced in a preliminary work~\cite{decrop2026analyzing}. Thus, the current contribution extends the idea according to the following:

\begin{itemize}
    \item The preliminary work initially presented 24 status code usage rules. In the current work, these rules were refined, extended, and validated against additional sources~\cite{masse2011rest}. There is now a total of \nbRules{} status code usage rules.
    \item The preliminary evaluation contained a limited dataset of 60 REST API specifications. In the current work, we greatly extend the dataset, now comprising a total of 2,625 distinct REST API specifications spanning across three state-of-the-art benchmarks~\cite{apisguru2026apis,decrop2025public,sahin2025wfc}.
    \item The current work improves \statictool{} with additional rules, a better implementation structure, and more complete specifications/documentation. Moreover, we also address status code misuses at a dynamic level with \evo, which was not a subject of the preliminary work.
\end{itemize}

We provide the background and related work in \Cref{sec:background}. Next, we describe the approaches (for \statictool{} and the extension of \evo) in \Cref{sec:approach}. Then, \Cref{sec:evaluation} presents our evaluation, while \Cref{sec:discussion} discusses misuse implications and provides recommendations for API clients and testers. \Cref{sec:threats} presents the threats to validity, and \Cref{sec:conclusion} wraps up the paper with a conclusion and future work.
\section{Background and Related Work}
\label{sec:background}

\subsection{REST APIs and HTTP}

REpresentational State Transfer (REST)~\cite{fielding2000architectural} is an architectural style providing a set of design principles for building scalable, loosely coupled, and interoperable web services. Such principles include stateless communication, where client requests must contain all necessary information for server processing. REST Application Programming Interfaces (APIs) use the HyperText Transfer Protocol (HTTP)~\cite{fielding2022rfc} to perform CRUD (Create, Read, Update, Delete) operations on data resources identified by URIs. In a HTTP request, paths (e.g., \texttt{/users}) and query parameters (e.g., \texttt{name=john}) describe what the API server should do. APIs implementing or extending REST are termed RESTful~\cite{richardson2013restful}.

\subsection{HTTP Methods and Requests}

HTTP implements various methods to indicate the purpose of requests. A total of nine methods are defined, however only five of them are frequently used:

\begin{itemize}
    \item \texttt{GET} is used to request a representation of data from the server.
    \item \texttt{POST} is used to send data to the server, often to create new data.
    \item \texttt{PUT} is used to either create or update data on the server, depending on its presence.
    \item \texttt{PATCH} is used to update data on the server.
    \item \texttt{DELETE} is used to delete data from the server.
\end{itemize}

\texttt{POST}, \texttt{PUT}, and \texttt{PATCH} methods usually contain a payload, consisting of the data to be created or updated. For instance, a client could send the request \texttt{POST /users} with a JSON payload \texttt{\{"name": "John"\}} to create a user named John on the server. Later, the client could send the request \texttt{GET /users?name=John} to retrieve the data of the newly created user. Additionally, the JSON format is widely adopted for representing response and payload data.

\begin{figure}[t]
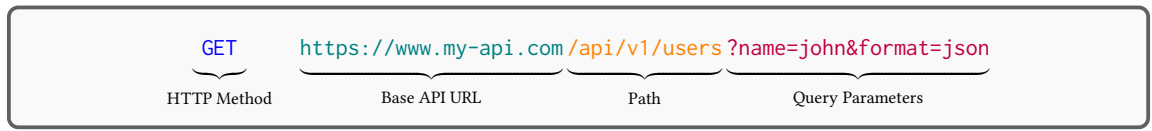

    \begin{tcolorbox}[colback=gray!5!white,colframe=gray!75!black]
      \[ \underbrace{\textcolor{blue}{\texttt{GET} \vphantom{q}}}_{\text{HTTP Method}} \quad \underbrace{\textcolor{teal}{\texttt{https://www.my-api.com} \vphantom{q}}}_{\text{Base API URL}} \underbrace{\textcolor{orange}{\texttt{/api/v1/users} \vphantom{q}}}_{\text{Path}} \underbrace{\textcolor{purple}{\texttt{?name=john\&format=json} \vphantom{q}}}_{\text{Query Parameters}} \]
    \end{tcolorbox}
    \caption{\label{fig:request-example} Structure of a typical HTTP GET request, for requesting data from a REST API server.}
    \Description{Figure of a structure of a typical HTTP GET request, for requesting data from a REST API server.}
\end{figure}

Combining an HTTP method with a path generates a REST API HTTP request. Depending on the desired operation, requests may contain payloads and parameters (for path, query, and/or header). \Cref{fig:request-example} illustrates an example structure for a typical \texttt{GET} request.

\subsection{HTTP Status Codes}
\label{subsec:http-status-codes}

HTTP also defines a set of status codes to communicate the outcome of client-server interactions. Status codes are grouped into five ranges: informational (\texttt{1xx}), success (\texttt{2xx}), redirection (\texttt{3xx}), client error (\texttt{4xx}), and server error (\texttt{5xx}). A status code is structured as a three-digit identifier, where the first digit indicates the range and the next two digits indicate the exact condition within that range. For instance, the status code \texttt{200 OK} indicates a successful request, \texttt{404 Not Found} indicates that a resource was not found, and \texttt{500 Internal Server Error} indicates a server-side failure. Consequently, status codes provide clear feedback to clients, facilitating error handling and debugging practices.

\begin{figure}[t]
    \centering
    \includegraphics[width=0.8\linewidth]{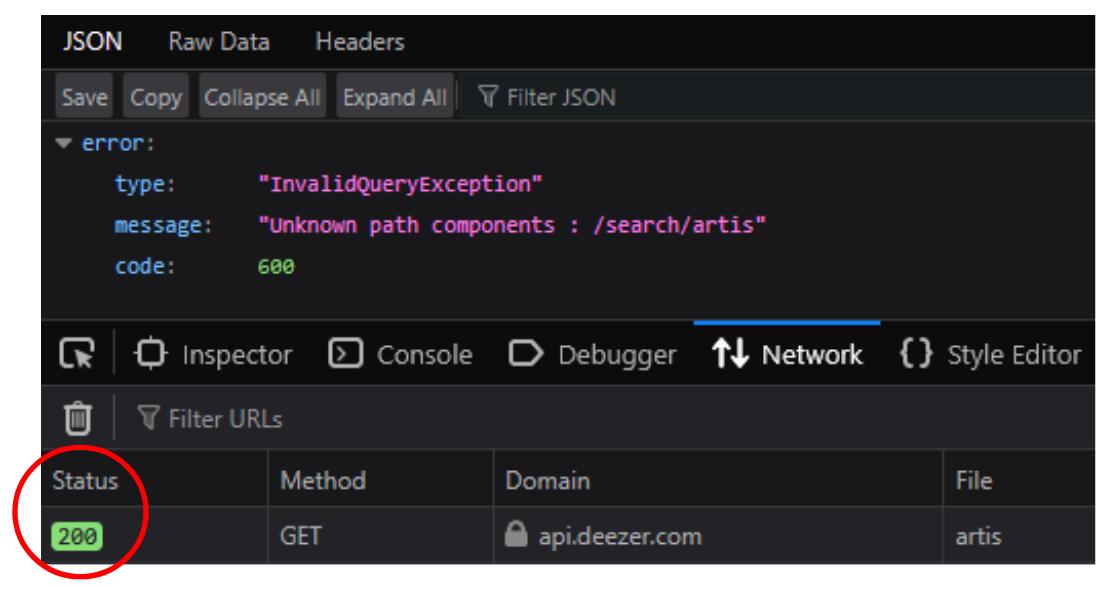}
    \caption{Example of a status code misuse.}
    \Description{Image of an example of a status code misuse.}
    \label{fig:status-code-misuse-example}
\end{figure}

Status codes are standardized by the Internet Engineering Task Force (IETF) and mainly defined in RFC 9110 (HTTP Semantics)~\cite{fielding2022rfc}. Each status code is described in terms of its general meaning and when it is appropriate to use. However, HTTP status code semantics are not enforced in practice for REST APIs, leading to potential misuses. \Cref{fig:status-code-misuse-example} illustrates an example of a status code misuse in the Deezer API~\cite{deezer2026deezer}. As shown, a \texttt{200 OK} status code is returned by the server, suggesting a success regarding the sent request. However, the response body contains an error message, indicating that the path does not exist. The error message also contains an API-defined \texttt{600} status code, which is non-standard as the \texttt{6xx} range is not defined in HTTP. In consequence, a client blindly trusting the status code would falsely believe that the request has succeeded, leading to a false positive.

\subsection{OpenAPI Specification}

The OpenAPI Specification (OAS)~\cite{linux2026openapi}, previously known as Swagger~\cite{smartbear2026swagger}, is a widely adopted industry standard to document REST APIs. The OAS standard is structured and machine-readable, written with either JSON or YAML file formats. Specifications can be organized into objects that describe the overall REST API (e.g., title, version, servers), resources and their paths, supported HTTP methods, request parameters, expected responses, and reusable components such as security schemes and data models.

OAS is widely used for REST API testing practices, as most state-of-the-art testing tools~\cite{golmohammadi2023testing} require an OAS file of the API under test as input. Moreover, editing tools such as the Swagger Editor~\cite{smartbear2026swaggereditor} can convert OAS files into human-readable documents.

\subsection{Related Work}

For related work, we focus on three main topics: REST API design conformance, HTTP semantics compliance, and REST API testing.

\subsubsection{REST API Design Conformance}

A baseline for REST API design conformance is the ``REST API Design Rulebook'' by Massé~\cite{masse2011rest}. The book provides a set of rules and guidelines to design and develop REST APIs, related to identifier design with URIs, interaction design with HTTP, metadata design, representation design, and client concerns.

More concrete applications include the work of Di Meglio, Pontillo, and Starace~\cite{di2025rest}, who analyzed RESTful design rule violations in 40 web apps built by students. Bogner et al. \cite{bogner2024restruler} implemented \textsc{RESTRuler}, a tool aimed at detecting design rule violations in OpenAPI descriptions. The tool implements 14 rules defined in Massé's work, and found 169,061 violations across 2,300 OpenAPI files, suggesting opportunities for improvement. Rodriguez et al. \cite{rodriguez2016rest} did a large scale analysis of compliance with REST API principles and best practices, revealing issues in the implementation and usage of REST APIs in terms of stability.

Existing tools~\cite{aptori2025openapi,bradburn2019openapi,specmantic2026specmantic} are capable of analyzing OpenAPI specification-implementation conformance by checking mismatches and/or undeclared data. Yet, such approaches require source code, do not cover fine-grained misuses, and/or do not focus specifically on HTTP status code semantics.

Our previous work~\cite{decrop2026analyzing} introduced \statictool{}, a static analysis tool capable of detecting status code misuses in OpenAPI specifications. The tool implements 24 status code usage rules, derived from official HTTP standards~\cite{fielding2022rfc}, REST API principles~\cite{fielding2000architectural}, and Massé's best practices~\cite{masse2011rest}. Our preliminary evaluation revealed over 17,767 status code misuses across 60 specifications, highlighting the importance of further improving the work.

\subsubsection{HTTP Semantics Compliance}

More broadly regarding HTTP semantics compliance, Rautenstrauch and Stock~\cite{rautenstrauch2024breaking} studied the landscape of HTTP conformance on the web, revealing security issues and a global lack of conformance in 9,990 web hosts. The authors highlights complicated specifications, missing negative feedback, and a lack of testing infrastructure as probable causes to this problem. Benhabbour, Attia, and Dacier~\cite{benhabbour2025http} verified if network middleboxes affect the conformance of HTTP traffic, revealing that 12 implementations were not compliant with RFCs due to ambiguities and leading to potential vulnerabilities.

\subsubsection{REST API Testing Tools}
There are several existing fuzzers that are able to dynamically generate HTTP requests to test REST APIs~\cite{golmohammadi2023testing}. 
These include for example
APIRL~\cite{foley2025apirl},
AutoRestTest~\cite{kim2025autoresttest},
EmRest~\cite{xu2025effective},
\evo~\cite{arcuri2025tool},
Morest~\cite{liu2022icse},
Nautilus~\cite{deng2023nautilus},
Schemathesis~\cite{hatfield2022deriving},
VoAPI2~\cite{du2024vulnerability}
and
WuppieFuzz~\cite{rooijakkers2025wuppiefuzz}.

To operate, these fuzzers need to know where the API is up and running (e.g., a URL address), and what is available to call.
This is typically expressed with a schema definition, like the OpenAPI format. 
Based on this information, and what returned in the responses of the HTTP calls, these fuzzers can use several different strategies to decide which endpoint to call, and what inputs to use (e.g., for query parameters and body payloads). 

These fuzzers can also automatically find faults based on returned HTTP 500 status code (Server Error)~\cite{marculescu2022faults}, or mismatches between the obtained HTTP responses and what declared possible in the OpenAPI schemas. 
Some REST API fuzzers can also do different kind of security attacks to detect security vulnerabilities (e.g., \cite{deng2023nautilus,du2024vulnerability,arcuri2025fuzzing}).

\subsection{Terminology}

In this paper, we use the term API as a shorthand for a REST API. When we mention a REST API specification, this refers to a documentation written in the OpenAPI Specification (OAS) standard. We also interchangeably use the terms ``status code misuse'' and ``status code usage rule violation''. Similarly, we interchangeably use the terms ``dataset'' and ``benchmark''.
\section{Approach}
\label{sec:approach}

\subsection{Static Analysis: SCOAS}
\label{subsec:scoas}

\begin{figure}[t]
    \centering
    \includegraphics[width=0.9\textwidth]{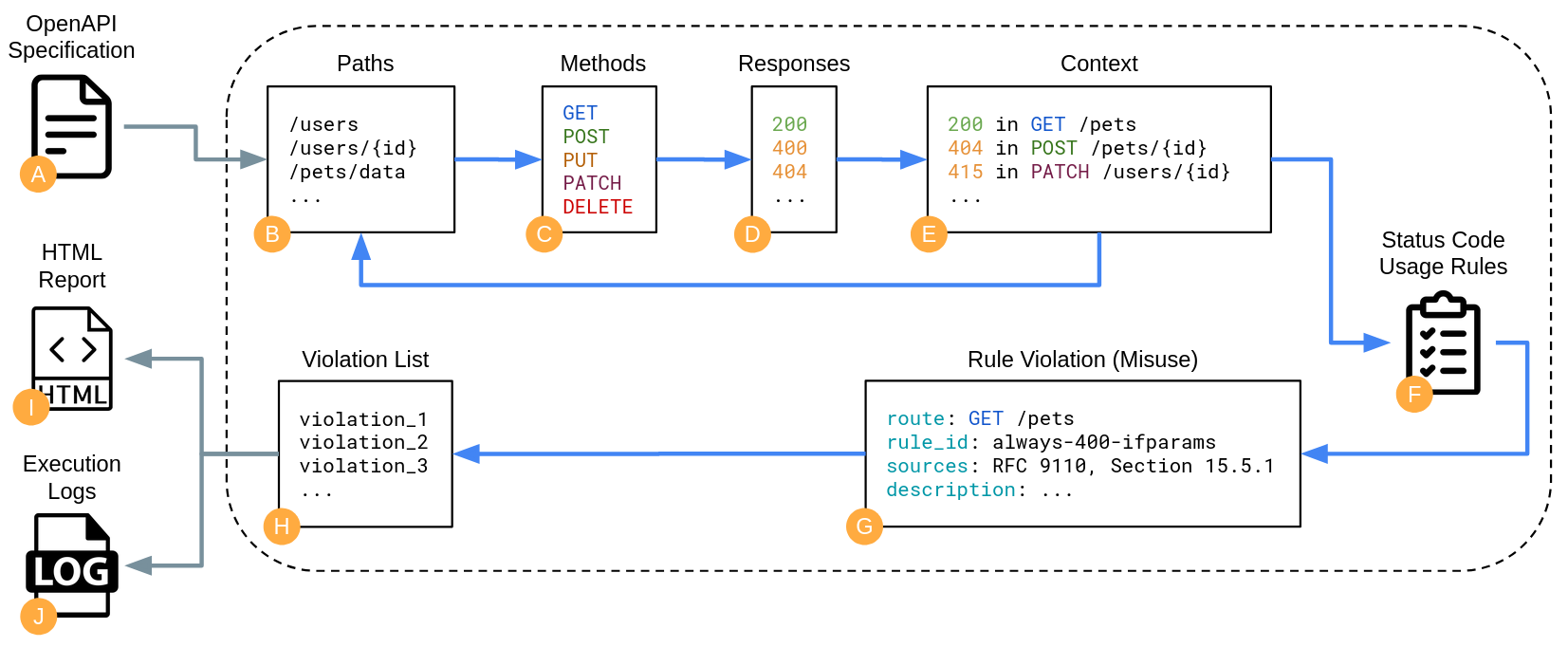}
    \caption{Overview of our tool SCOAS.}
    \Description{Image for: Overview of our tool SCOAS.}
    \label{fig:tool}
\end{figure}

To detect status code misuses statically, we developed SCOAS (Status Code Analysis in OpenAPI Specifications). \Cref{fig:tool} illustrates our tool, with orange circles \circled{X} representing each phase. First, an OAS file (JSON format) of a REST API is provided \circled{A}. Then, we iterate over all paths specified in the file \circled{B}. For each path, we iterate over all described methods \circled{C}. For each method in each path, we iterate over all described responses \circled{D}. Doing so allows us to generate a status code context \circled{E}, composed of a path, method, and response. This is required, as status code usage rules may rely on path data (e.g., path parameters such as \texttt{/\{id\}} for \texttt{404 Not Found} responses) and/or method data (e.g., \texttt{GET} methods require \texttt{200 OK} responses). When a status code context is found, we iterate over all defined status code usage rules \circled{F} (defined in \Cref{sec:evaluation}). We verify if rules are violated by performing a static analysis of the context against the rules. If it is the case, we generate a rule violation (misuse) containing the triggering context and related information \circled{G}. We append it to a list of violations \circled{H}. The process continues until all status code contexts have been analyzed. SCOAS stops when the iterations are completed (\circled{B} - \circled{E}). Upon ending, SCOAS issues an HTML report \circled{I} and execution logs \circled{J}.

\begin{figure}[t]
    \centering
    \includegraphics[width=0.9\textwidth]{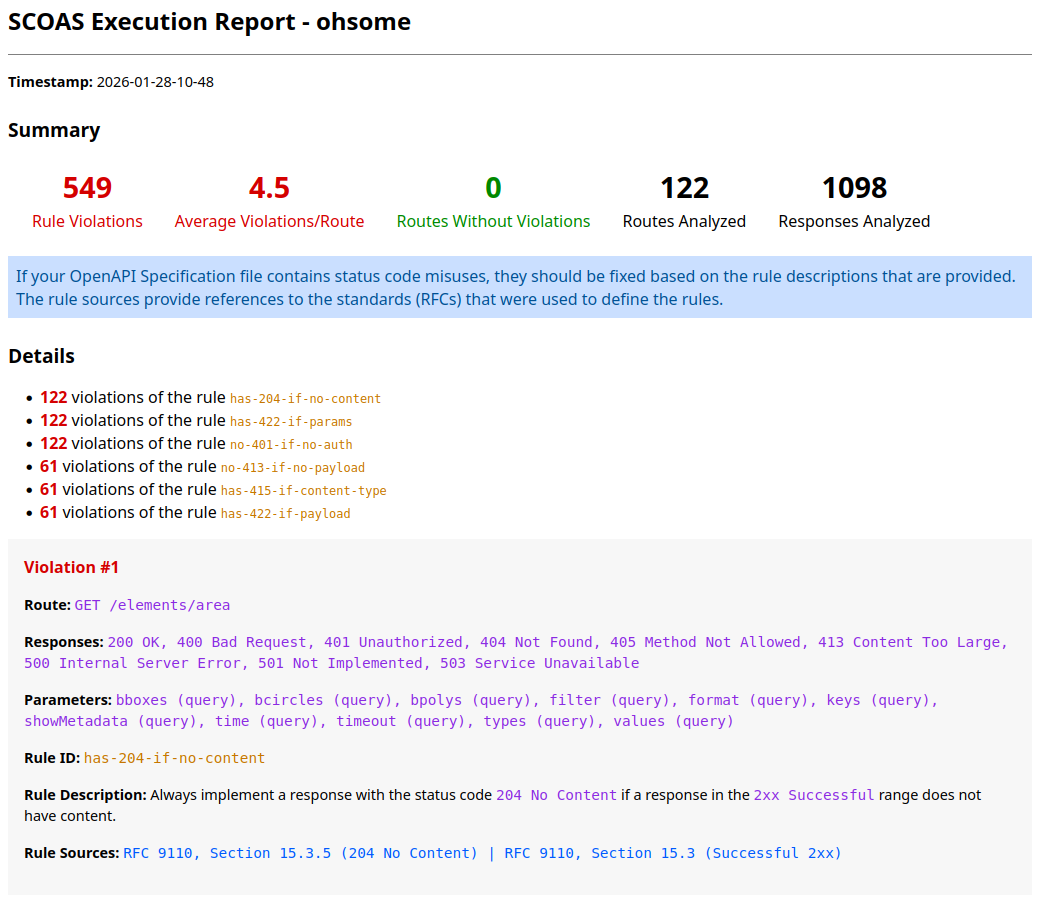}
    \caption{Example of an HTML report generated by SCOAS.}
    \Description{Image of: Example of an HTML report generated by SCOAS.}
    \label{fig:scoas-report-example}
\end{figure}

\Cref{fig:scoas-report-example} illustrates an example of an HTML report that the tool can generate. As shown, a summary is given with overall execution data (total number of rule violations, average violations per route, responses analyzed, etc.), followed by specific data for each reported violation. For each violation, the context (i.e., route, responses, and parameters) and rule information (i.e., identifier, description, and sources) are given. This allows API developers to better understand their violations and instructions to fix them.

\subsection{Dynamic Analysis: EvoMaster}
\label{subsec:dynamicanalysis}

HTTP semantics properties can  be \emph{dynamically} evaluated during the fuzzing process. 
Each time an HTTP call is made, a fuzzer can check the returned HTTP status code, and evaluate whether it is correct or not based on the context (e.g., used HTTP verb).
However, compared to \emph{static} analysis on the OpenAPI schemas (recall Section~\ref{subsec:scoas}) there are a few differences that need to be taken into account. 

First, dynamic fuzzing can show the presence of a problem, but not prove its absence. 
In other words, testing can find faults, but cannot guarantee that a software is fault-free. 
In practice, this means that, dynamically, we can only evaluate a \emph{subset} of oracles that can be checked statically. 
All oracles that are based on something that must be present in the schema would not be possible to evaluate dynamically.
For example, assume we want to make sure that each \texttt{GET} operation has at least one possible response that returns an HTTP \texttt{200 OK} status code. 
Even if left running for hours, a fuzzer could fail to generate the right input data to get such result, and only obtain user-error responses (e.g., a \texttt{400} or \texttt{404}). 
Stating something like there is a fault in the API because no response with a \texttt{200} was obtained would just be a false positive. 
On the other hand, if a \texttt{201 Created} response is returned in a \texttt{GET} request, the presence of that status code for that HTTP verb would be a clear HTTP semantics fault.

The OpenAPI schema might define some wrong status codes (e.g., as previously discussed, a \texttt{201} on a \texttt{GET} endpoint). 
This does not mean though that the API would ever return a response with such status code. 
So, \emph{static} analysis on the OpenAPI schema could detect a documentation fault that would never happen in practice for the clients of the API. 
In contrast, \emph{dynamic} analysis has the benefit that it can then create executable test cases to show that this could indeed happen in practice. 

A further advantage is that it can detect this kind of issue regardless of what specified in the OpenAPI schema. 
For example, an endpoint might not specify a \texttt{201} code for a \texttt{GET} (and so static analysis would not be able to report any issue). 
Still, if a \texttt{201} code is returned, dynamic analysis would point out the problem. 
The interesting point here is that, technically, a situation like this would represent two distinct faults.
First, a \texttt{201} is returned for a \texttt{GET} (HTTP Semantics Fault) and, second, a response is returned with a status code not declared in the schema (Schema Mismatch Fault). 

As long as there is no requirement in constructing specific sequences of HTTP calls, checking HTTP oracles related to status codes is relatively easy to implement, and to integrate to any existing fuzzer.  
In our case, for our empirical study, it took less than 100 lines of code to add and integrate all our defined HTTP oracles to the state-of-the-art fuzzer \evo~\cite{arcuri2025tool}. 
\section{Evaluation}
\label{sec:evaluation}

\subsection{Research Questions}

For our evaluation, we formulate the following research questions:

\begin{itemize}
    \item \textbf{RQ.1:} \textit{Which HTTP status codes are most relevant for REST APIs?}
    \item \textbf{RQ.2:} \textit{What status code usage rules may be established for REST APIs?}
    \item \textbf{RQ.3:} \textit{How effective is static analysis at detecting status code misuses in REST APIs?}
    \item \textbf{RQ.4:} \textit{How effective is dynamic analysis at detecting status code misuses in REST APIs?}
\end{itemize}

\subsection{Experimental Setup}

\subsubsection{Dataset and Benchmark}
\label{subsubsec:dataset}

For our evaluation, we formed a comprehensive dataset of REST APIs from three different state-of-the-art benchmarks:

\begin{enumerate}
    \item \textbf{APIs.guru~\cite{apisguru2026apis}:} A dataset containing over 2,500 public REST API definitions, provided in the OpenAPI Specification (OAS) format. The documented APIs are open-source, public, and updated on a weekly basis. The evaluation charts for this dataset are colored in blue.
    
    \item \textbf{Public REST API Benchmark (PRAB)~\cite{decrop2025public}:} A dataset of 60 REST APIs which are commonly used in the relevant academic literature (i.e., works related to REST APIs). The dataset provides OpenAPI specifications and structural characteristics (e.g., routes, query parameters, HTTP methods, authentication) for the APIs. The evaluation charts for this dataset are colored in red.

    \item \textbf{Web Fuzzing Dataset (WFD)~\cite{sahin2025wfc}:} A dataset of 36 open-source REST APIs, providing necessary information (i.e., source code, schemas, Docker files and tool support) for external tools (e.g., fuzzer) to run experiments. The evaluation charts for this dataset are colored in green.
\end{enumerate}

The three datasets are used for RQ.1, RQ.2, and RQ.3, as they all provide uncomplicated access to OpenAPI specifications of REST APIs. However, as RQ.4 requires an advanced setup for dynamic analysis, solely WFD is used for the related evaluation, being tailored for such purpose.

As these datasets often select public REST APIs, it is possible that some duplicates may appear. However, as evaluation results are analyzed individually in between the datasets, duplicates may be kept as removing some APIs from a certain dataset might skew the results. Additionally, APIs may not exhibit the same version across different datasets, further hindering duplicate detection.

\subsubsection{Environment}

We ran our evaluations using the following environments and configurations:

\begin{itemize}
    \item For RQ.1, RQ.2, and RQ.3 (i.e., status code distribution, usage rules, and static analysis), the experiments were executed on a Lenovo Thinkbook 14s Yoga ITL 20WE laptop with the following specifications: Processor Intel(R) Core(R) i5-1135G7 @ 2.40GHz × 4; RAM 16GB; Operating System 64-bit Linux Mint 22.3 Cinnamon. As the approach is static and entirely offline, a single execution was sufficient for the evaluation results. Yet, to avoid any external issues (e.g., hardware-related or console crashes), multiple executions were carried out, all yielding identical results.

    \item For RQ.4, the experiments with \evo were executed on an HP Z6 G4 Workstation with the following specifications: Processor Intel(R) Xeon(R) Gold 6240R CPU @ 2.40GHz 2.39GHz; RAM 192GB; Operating System 64-bit Windows 11. Experiments were run in parallel, 15 at a time.
\end{itemize}

\subsection{RQ.1: Relevant Status Codes for REST APIs}
\label{subsec:rq1}

As mentioned in \Cref{subsec:http-status-codes}, HTTP status codes span across five ranges: informational (\texttt{1xx}), success (\texttt{2xx}), redirection (\texttt{3xx}), client error (\texttt{4xx}), and server error (\texttt{5xx}). From these five ranges, a total of 500 status codes can theoretically be utilized. Yet, not all status codes are relevant in the context of REST APIs and their plausible interactions. For instance, we may exclude the informational (\texttt{1xx}) range, which is intended for low-level HTTP connection control (e.g., protocol switching)~\cite{fielding2022rfc}. As REST APIs operate at the application layer~\cite{fielding2000architectural}, there is no connection negotiation and thus \texttt{1xx} status codes offer no semantic value in REST APIs. Similarly, the redirection (\texttt{3xx}) range is rare in REST APIs as it is primarily designed for web navigation and client-side redirection (e.g., browsers following links). In contrast, REST APIs typically return direct responses without requiring the client to perform additional request hops. The remaining three ranges are widely used in REST APIs:

\begin{itemize}
    \item The success (\texttt{2xx}) range is used to indicate successful request handling, comprising specific codes which may be mapped to CRUD operations (e.g., \texttt{200 OK} for a successful retrieval, \texttt{201 Created} for a successful creation).

    \item The client error (\texttt{4xx}) range is used to indicate problems with client requests (e.g., invalid parameter syntax, invalid headers, forbidden access), which enables clear and actionable error reporting.

    \item The server error (\texttt{5xx}) range is used to signal server-side failures (i.e., a request may not be processed due to backend issues). This range is especially useful in REST APIs to distinguish client and server responsibilities.
\end{itemize}

While these ranges are relevant in REST APIs, not all of their individual status codes are commonly used in practice. For instance, the \texttt{205 Reset Content} status code indicates that the server has successfully processed the request and instructs the client to reset its document view, while returning no content in the response. This behavior may be observed for content that supports data entry (e.g., web forms), which is not usually the case for REST API content.

In consequence, we extracted relevant status codes based on their usage in real-world REST APIs. To do so, we analyzed the OpenAPI specifications of the APIs contained in the evaluation dataset. The utilized status codes may be found by iterating over the \texttt{responses} field, contained in the routes (paths + HTTP methods) of the API documentation. We define the prevalence of a status code $c$ according to the following formula:

\begin{equation*}
    \mathrm{Prevalence}(c) = \frac{\#\text{APIs where } c \text{ appears at least once}}{\#\text{APIs in the dataset}}
\end{equation*}

\vspace{0.25cm}

\begin{figure}[t]
    \centering
    \includegraphics[width=0.9\linewidth]{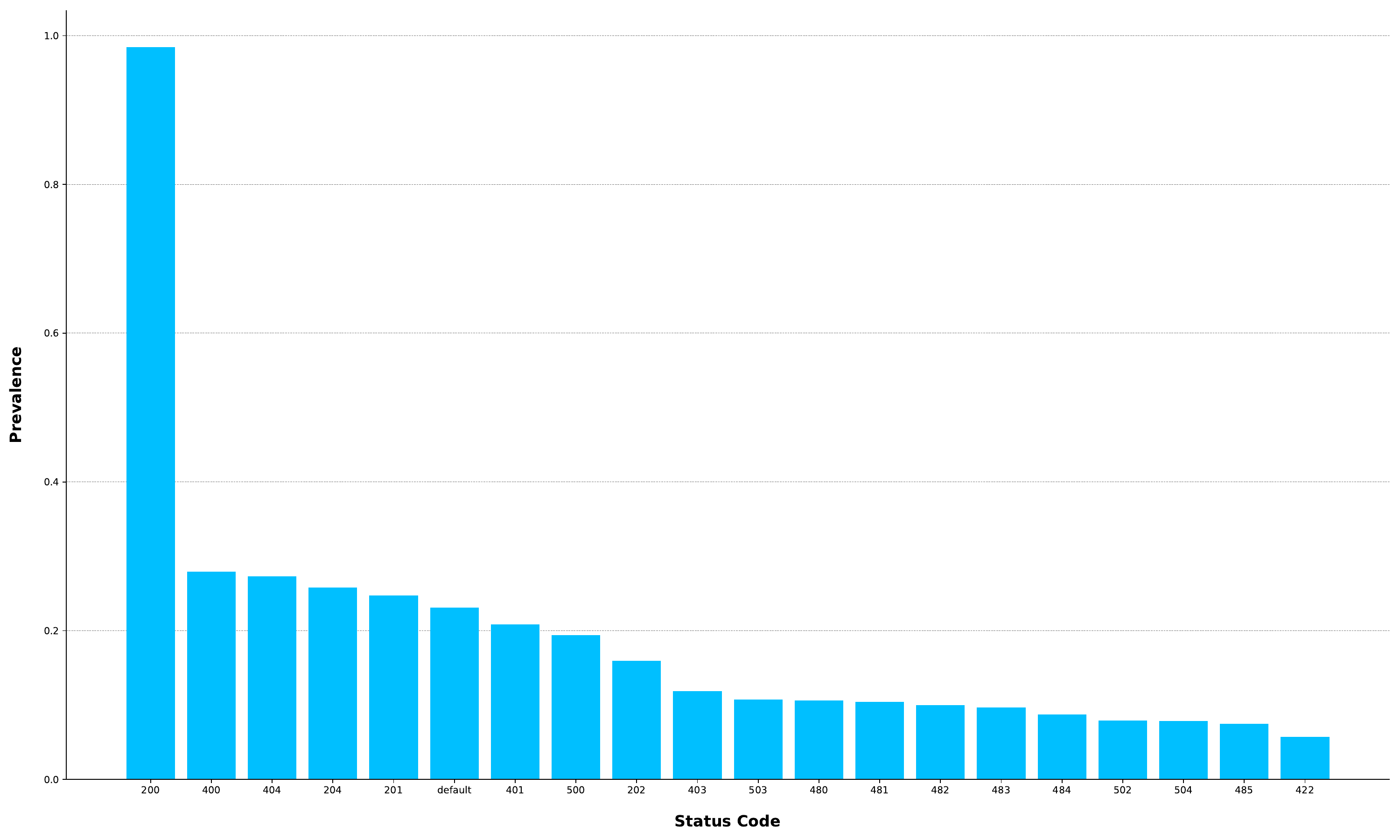}
    \caption{Distribution of status code prevalence for the APIs.guru dataset, for the 20 most prevalent status codes.}
    \Description{Image of the distribution of status code prevalence for the APIs.guru dataset, for the 20 most prevalent status codes.}
    \label{fig:chart-sc-distrib-apisguru}
\end{figure}

\begin{figure}[t]
    \centering
    \includegraphics[width=0.9\linewidth]{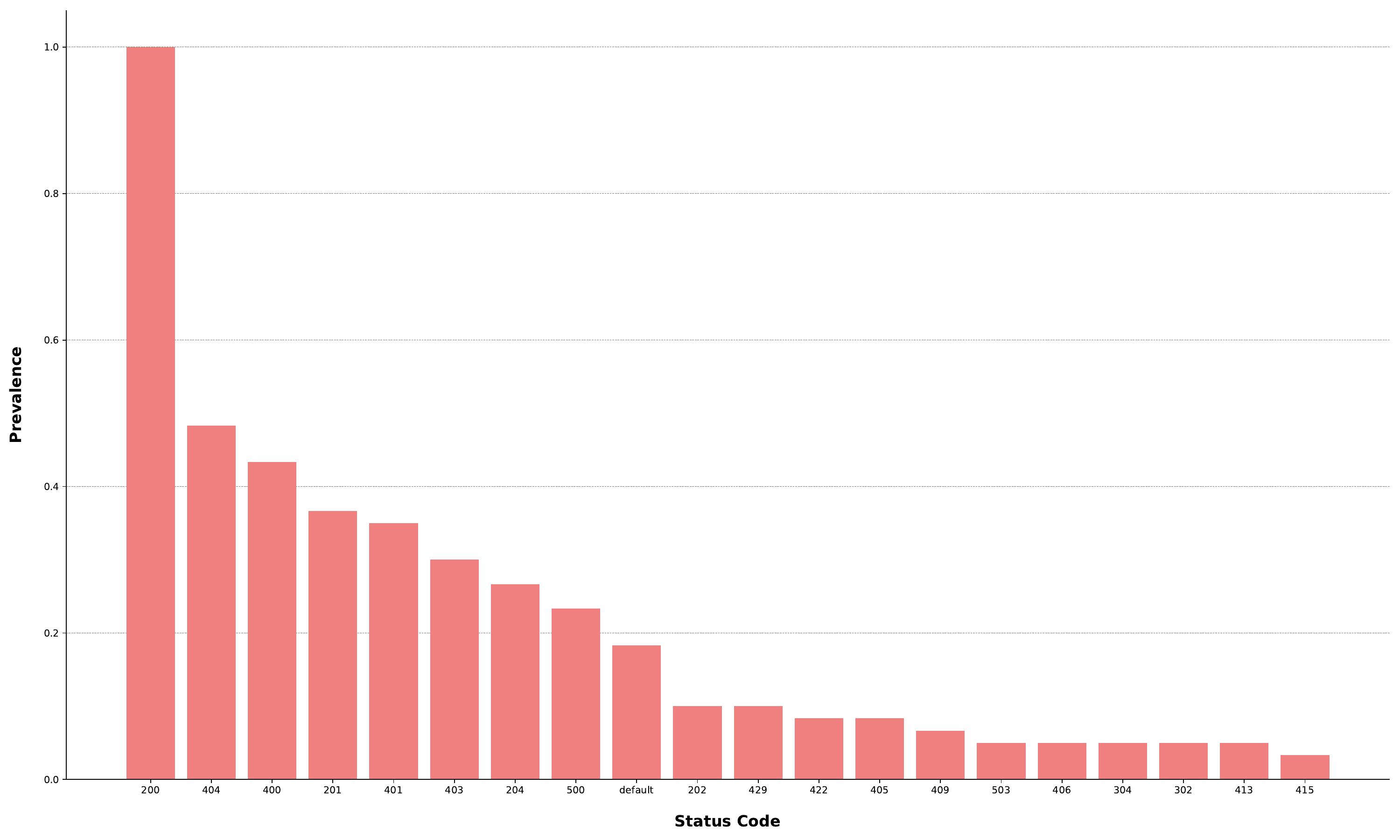}
    \caption{Distribution of status code prevalence for the PRAB dataset, for the 20 most prevalent status codes.}
    \Description{Image of the distribution of status code prevalence for the PRAB dataset, for the 20 most prevalent status codes.}
    \label{fig:chart-sc-distrib-prab}
\end{figure}

\begin{figure}[t]
    \centering
    \includegraphics[width=0.9\linewidth]{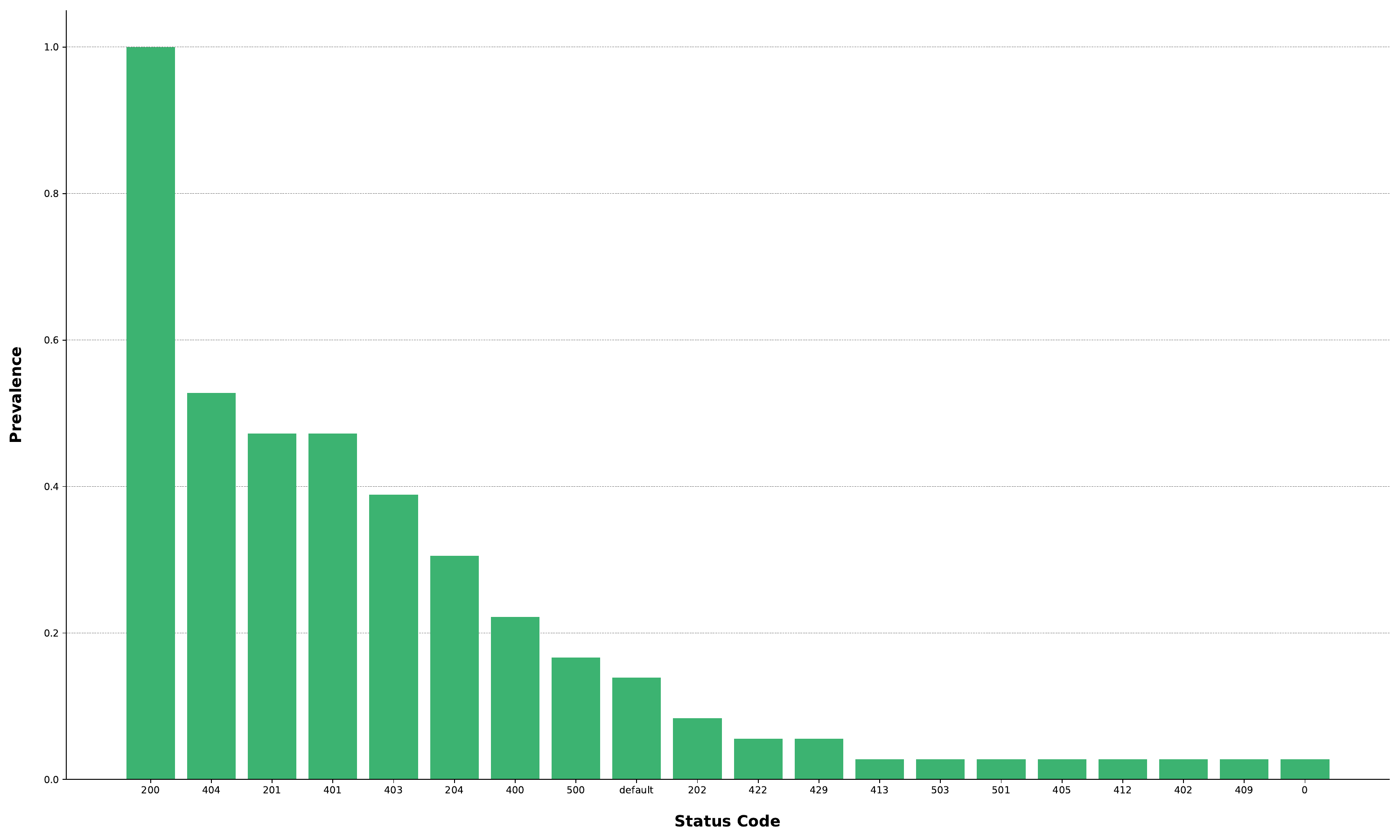}
    \caption{Distribution of status code prevalence for the WFD dataset, for the 20 most prevalent status codes.}
    \Description{Image of the distribution of status code prevalence for the WFD dataset, for the 20 most prevalent status codes.}
    \label{fig:chart-sc-distrib-wfd}
\end{figure}

Doing so allows us to measure how widely a status code is adopted across APIs, independent of how often it is used within each API and thus avoiding size bias. We report the results for the three evaluation datasets, and for the 20 most prevalent status codes. \Cref{fig:chart-sc-distrib-apisguru} illustrates the distribution for the APIs.guru dataset, \Cref{fig:chart-sc-distrib-prab} illustrates the distribution for the PRAB dataset, and \Cref{fig:chart-sc-distrib-wfd} illustrates the distribution for the WFD dataset. Overall, we analyzed and found:

\begin{itemize}
    \item 2,529 OAS files with 352,900 responses and 127 distinct status codes for the APIs.guru dataset.
    
    \item 60 OAS files with 12,074 responses and 35 distinct status codes for the PRAB dataset.

    \item 36 OAS files with 4,606 responses and 20 distinct status codes for the WFD dataset.
\end{itemize}

As shown, the most prevalent status code is \texttt{200 OK}, displaying a prevalence of 1.00 for the PRAB and WFD datasets, and a prevalence of 0.98 for the APIs.guru dataset. These results align with the well-established popularity of the \texttt{200 OK} status code in REST APIs, which remains the most frequently used response for successful HTTP requests across many services.

Next, we observe that many status codes from the client error (\texttt{4xx}) range are prevalent: \texttt{400 Bad Request}, \texttt{401 Unauthorized}, \texttt{403 Forbidden}, and \texttt{404 Not Found}. These results suggest that REST APIs commonly employ client errors for invalid request syntax, authorization issues, access issues, and invalid paths respectively. These results also suggest that many APIs employ authentication mechanisms, due to the prevalence of \texttt{401} and \texttt{403} codes (especially in the WFD dataset).

The \texttt{500 Internal Server Error} status code is also prevalent, appearing in the results for all datasets. Similarly to the \texttt{200 OK} status code, it is prone to misuses as it is often used for generic error response, regardless of client/server concerns. Moreover, \texttt{default} appears in the results for all datasets. While this keyword is not a standard HTTP status code, it is defined in the OpenAPI Specification standard and is used to collectively describe errors that have the same response structure. Yet, it is widely adopted in practice and can be considered as valid when documenting REST APIs through this format. Nonetheless, specific status codes should be used as much as possible to describe responses in a more precise way.

Furthermore, we observe that many non-standard and unassigned (i.e., syntactically valid yet without any semantic meaning) status codes appear in the results. For instance, the status codes \texttt{480} to \texttt{485} appear in the distribution of the APIs.guru dataset. However, such codes are officially unassigned and thus provide no information for API clients relying on HTTP semantics. In the PRAB and WFD datasets, we observe the presence of a non-standard status code \texttt{0}. This status code is wrong, however it is used in some libraries to represent OAS \texttt{default}, when only digits are supported for status codes and not strings. Thus, the digit \texttt{0} is used in such cases as it is unassigned in HTTP semantics. In lesser numbers, we also find \texttt{5XX} in the APIs.guru and PRAB datasets, which is a range and not an exact status code. These results highlight that REST APIs may use non-standard or unassigned status codes, hindering client-side understandability.

\begin{tcolorbox}
    \textbf{RQ.1 Summary}: After analyzing 369,580 responses in the 2,625 OpenAPI specifications of the datasets, we reported the prevalence of 128 distinct status codes. We found that REST APIs commonly employ status codes from the success (\texttt{2xx}) and client error (\texttt{4xx}) ranges, aligning with their use in typical client-server interactions. We identified multiple occurrences of security-related status codes (\texttt{401} and \texttt{403}), suggesting that real-world REST APIs employ authentication mechanisms in practice. However, we observed uses of non-standard and unassigned status codes, hindering reliance on HTTP semantics for clients.
\end{tcolorbox}

\subsection{RQ.2: Defining Status Code Usage Rules}
\label{subsec:rq2}

In order to identify status code misuses in REST APIs, we first need to define a set of status code usage rules. Then, we can verify if the APIs comply to such rules based on specific context (e.g., routes, parameters, HTTP methods, and responses). 

First, we read and analyzed the content of the ``Hypertext Transfer Protocol (HTTP) Status Code Registry'' defined by the Internet Assigned Numbers Authority (IANA)~\cite{iana2025hypertext}. It is widely considered as the official registry for status codes. The registry details all status codes along with their official RFC (Request For Comments) Internet Standards, describing each status code with additional information and details. Most status code references redirect to RFC 9110~\cite{fielding2022rfc} (entitled ``HTTP Semantics''), which is the latest and most up-to-date RFC for status codes. For rules related to the \texttt{PATCH} HTTP method, RFC 5789~\cite{dusseault2010rfc} (entitled ``PATCH Method for HTTP'') is also utilized as a reference. We identified statement verbs (e.g., ``must'', ``must not'', etc.) and mapped them to adequate rules for REST API responses.

Second, we also defined rules based on Massé's book on REST API design~\cite{masse2011rest}. Indeed, Chapter 3 of the book (entitled ``Interaction Design with HTTP'') defines a set of rules related to HTTP method and status code usage for REST APIs. For instance, Massé states that ``\texttt{404 Not Found} must be used when a client's URI cannot be mapped to a resource'' (Chapter 3, page 31). In this case, we adapted to rule so that the presence of the status code could be detected in OAS files, if it is relevant for a given context. This rule is a formalization of RFC 9110 Section 15.5.5 (entitled ``404 Not Found''), which details use cases for this status code. In consequence, we compared the rules defined by Massé with the relevant RFC sections to cross-check their validity.

Third, we considered REST API principles~\cite{fielding2000architectural} (e.g., GET HTTP method does not allow creation) and the status code distribution identified in RQ.1 (cf. \Cref{subsec:rq1}). The distribution allowed us to focus on defining status code usage rules for the success (\texttt{2xx}) and client error (\texttt{4xx}) ranges. Some rules for the redirection (\texttt{3xx}) and server error (\texttt{5xx}) ranges were also defined. We did not consider the informational (\texttt{1xx}) range, as RQ.1 demonstrated that it is not tailored for usual REST API interactions.

After an exhaustive analysis of these three external resources, we were able to define a total of \nbRules{} unique status code usage rules for REST APIs. Table \ref{tab:status-code-rules} details the rules, which are represented by a unique \textbf{Identifier}, a \textbf{Description}, and \textbf{Sources} (corresponding to the official RFCs and/or Massé's book pages that were used to define the rule). The usage rules may be categorized into two distinct groups: rules that verify if a mandatory status code is implemented (\texttt{has}), and rules that verify if a non-applicable status code is implemented (\texttt{no}).

\begingroup
\renewcommand{\arraystretch}{1.3}
\begin{table}
    \centering
    \caption{Defined status code usage rules for REST APIs. \textbf{I} = Implement, \textbf{NI} = Never Implement.}
    \label{tab:status-code-rules}
    \scriptsize
    \begin{tabular}{l p{5.2cm} p{3.8cm}}
        \toprule

        \textbf{Identifier} & \textbf{Description} & \textbf{Sources}\\

        \midrule

        \texttt{has-200-if-get} & \textbf{I} \texttt{200 OK} in a \texttt{GET} method. & RFC 9110 (Sec. 9, 15), Massé (p. 24, 28) \\
        
        \rowcolor{gray!30} \texttt{has-200-or-201-or-204-if-post} & \textbf{I} \texttt{200 OK}, \texttt{201 Created}, or \texttt{204 No Content} in a \texttt{POST} method. & RFC 9110 (Sec. 9, 15) \\
        
        \texttt{has-200-or-201-or-204-if-put} & \textbf{I} \texttt{200 OK}, \texttt{201 Created}, or \texttt{204 No Content} in a \texttt{PUT} method. & RFC 9110 (Sec. 9, 15) \\
        
        \rowcolor{gray!30} \texttt{has-200-or-204-if-delete} & \textbf{I} \texttt{200 OK} or \texttt{204 No Content} in a \texttt{DELETE} method. & RFC 9110 (Sec. 9, 15) \\
        
        \texttt{has-200-or-204-if-patch} & \textbf{I} \texttt{200 OK} or \texttt{204 No Content} in a \texttt{PATCH} method. & RFC 9110 (Sec. 9, 15), RFC 5789 (Sec. 2) \\
        
        \rowcolor{gray!30} \texttt{has-204-if-no-content} & \textbf{I} \texttt{204 No Content} if a response in the \texttt{2xx Successful} range does not have content. & RFC 9110 (Sec. 3, 15), Massé (p. 29) \\
        
        \texttt{has-400-if-params} & \textbf{I} \texttt{400 Bad Request} if the method contains parameters (in case of invalid syntax). & RFC 9110 (Sec. 5, 15) \\
        
        \rowcolor{gray!30} \texttt{has-400-if-payload} & \textbf{I} \texttt{400 Bad Request} if the method contains a payload (in case of invalid syntax). & RFC 9110 (Sec. 3, 15) \\
        
        \texttt{has-404-if-path} & \textbf{I} \texttt{404 Not Found} if the method contains path parameters. & RFC 9110 (Sec. 5, 15), Massé (p. 31) \\
        
        \rowcolor{gray!30} \texttt{has-406-if-accept} & \textbf{I} \texttt{406 Not Acceptable} in case the sever does not support the \texttt{Accept} header. & RFC 9110 (Sec. 12, 15), Massé (p. 32) \\
        
        \texttt{has-413-if-content-length} & \textbf{I} \texttt{413 Content Too Large} in case the server does not support the \texttt{Content-Length} header. & RFC 9110 (Sec. 8, 15) \\
        
        \rowcolor{gray!30} \texttt{has-415-if-content-type} & \textbf{I} \texttt{415 Unsupported Media Type} in case the server does not support the \texttt{Content-Type} header. & RFC 9110 (Sec. 8, 15), Massé (p. 32) \\
        
        \texttt{has-422-if-params} & \textbf{I} \texttt{422 Unprocessable Content} if the method contains parameters (in case of invalid semantics). & RFC 9110 (Sec. 5, 15) \\
        
        \rowcolor{gray!30} \texttt{has-422-if-payload} & \textbf{I} \texttt{422 Unprocessable Content} if the method contains a payload (in case of invalid semantics). & RFC 9110 (Sec. 3, 15) \\
        
        \texttt{no-200-if-error} & \textbf{NI} \texttt{200 OK} if the response content describes an error. & RFC 9110 (Sec. 3, 15), Massé (p. 28) \\
        
        \rowcolor{gray!30} \texttt{no-201-if-delete} & \textbf{NI} \texttt{201 Created} in a \texttt{DELETE} method (as it can never create data). & RFC 9110 (Sec. 9, 15) \\
        
        \texttt{no-201-if-get} & \textbf{NI} \texttt{201 Created} in a \texttt{GET} method (as it can never create data). & RFC 9110 (Sec. 9, 15) \\
        
        \rowcolor{gray!30} \texttt{no-201-if-patch} & \textbf{NI} \texttt{201 Created} in a \texttt{PATCH} method (as it can never create data). & RFC 9110 (Sec. 9, 15), RFC 5789 (Sec. 2) \\
        
        \texttt{no-204-if-content} & \textbf{NI} \texttt{204 No Content} if its content is not empty. & RFC 9110 (Sec. 3, 15), Massé (p. 29) \\

        \rowcolor{gray!30} \texttt{no-205-if-content} & \textbf{NI} \texttt{205 Reset Content} if its content is not empty. & RFC 9110 (Sec. 3, 15) \\

        \texttt{no-304-if-no-get-or-head} & \textbf{NI} \texttt{304 Not Modified} if the method is not \texttt{GET} or \texttt{HEAD}. & RFC 9110 (Sec. 9, 15) \\
        
        \rowcolor{gray!30} \texttt{no-401-if-no-auth} & \textbf{NI} \texttt{401 Unauthorized} if the specification does not contains an authentication mechanism. & RFC 9110 (Sec. 15), Massé (p. 31) \\

        \texttt{no-401-if-no-www-authenticate} & \textbf{NI} \texttt{401 Unauthorized} if it does not return a \texttt{WWW-Authenticate} header. & RFC 9110 (Sec. 11, 15) \\
        
        \rowcolor{gray!30} \texttt{no-403-if-no-401} & \textbf{NI} \texttt{403 Forbidden} if the method does not implement \texttt{401 Unauthorized}. & RFC 9110 (Sec. 15), Massé (p. 31) \\

        \texttt{no-405-if-no-allow} & \textbf{NI} \texttt{405 Method Not Allowed} if it does not return an \texttt{Allow} header. & RFC 9110 (Sec. 10, 15), Massé (p. 31) \\
        
        \rowcolor{gray!30} \texttt{no-413-if-no-payload} & \textbf{NI} \texttt{413 Content Too Large} if the method does not contain a payload. & RFC 9110 (Sec. 3, 15) \\
        
        \texttt{no-415-if-no-payload} & \textbf{NI} \texttt{415 Unsupported Media Type} if the method does not contain a payload. & RFC 9110 (Sec. 3, 15), Massé (p. 32) \\

        \rowcolor{gray!30} \texttt{no-426-if-no-upgrade} & \textbf{NI} \texttt{426 Upgrade Required} if it does not return an \texttt{Upgrade} header. & RFC 9110 (Sec. 7, 15) \\

        \texttt{no-501-if-implemented} & \textbf{NI} \texttt{501 Not Implemented} if the method is implemented. & RFC 9110 (Sec. 15) \\
        
        \rowcolor{gray!30} \texttt{no-non-standard-codes} & \textbf{NI} responses with non-standard status codes. & RFC 9110 (Sec. 15), OpenAPI doc. \\

        \bottomrule
    \end{tabular}
\end{table}
\endgroup

\begin{tcolorbox}
    \textbf{RQ.2 Summary}: After an exhaustive analysis of HTTP standards and REST API principles, we were able to define a total of \nbRules{} status code usage rules which are relevant to REST APIs. The rules may be used to find two types of status code misuses: missing status codes that should be mandatory, and implemented status codes that are not applicable to a given context.
\end{tcolorbox}

\subsection{RQ.3: Static Analysis of Status Code Misuses in REST APIs}

To evaluate the static analysis of status code misuses in REST APIs, we utilized the tool \textsc{SCOAS} (cf. \Cref{subsec:scoas}). The objective is to report violations of status code usage rules stemming from the OpenAPI specifications of the three datasets and provide additional insight regarding their distribution, frequency, and impact. We execute the tool on each dataset separately. Doing so allows us to identify if similar misuse patterns appear in the results of each dataset.

\subsubsection{Status Code Misuse Prevalence}

We define the prevalence of a status code misuse $m$ (i.e., a status code usage rule violation) according to the following formula:

\begin{equation*}
    \mathrm{Prevalence}(m) = \frac{\#\text{APIs where } m \text{ appears at least once}}{\#\text{APIs in the dataset}}
\end{equation*}

\vspace{0.25cm}

\begin{figure}[t]
    \centering
    \includegraphics[width=0.9\linewidth]{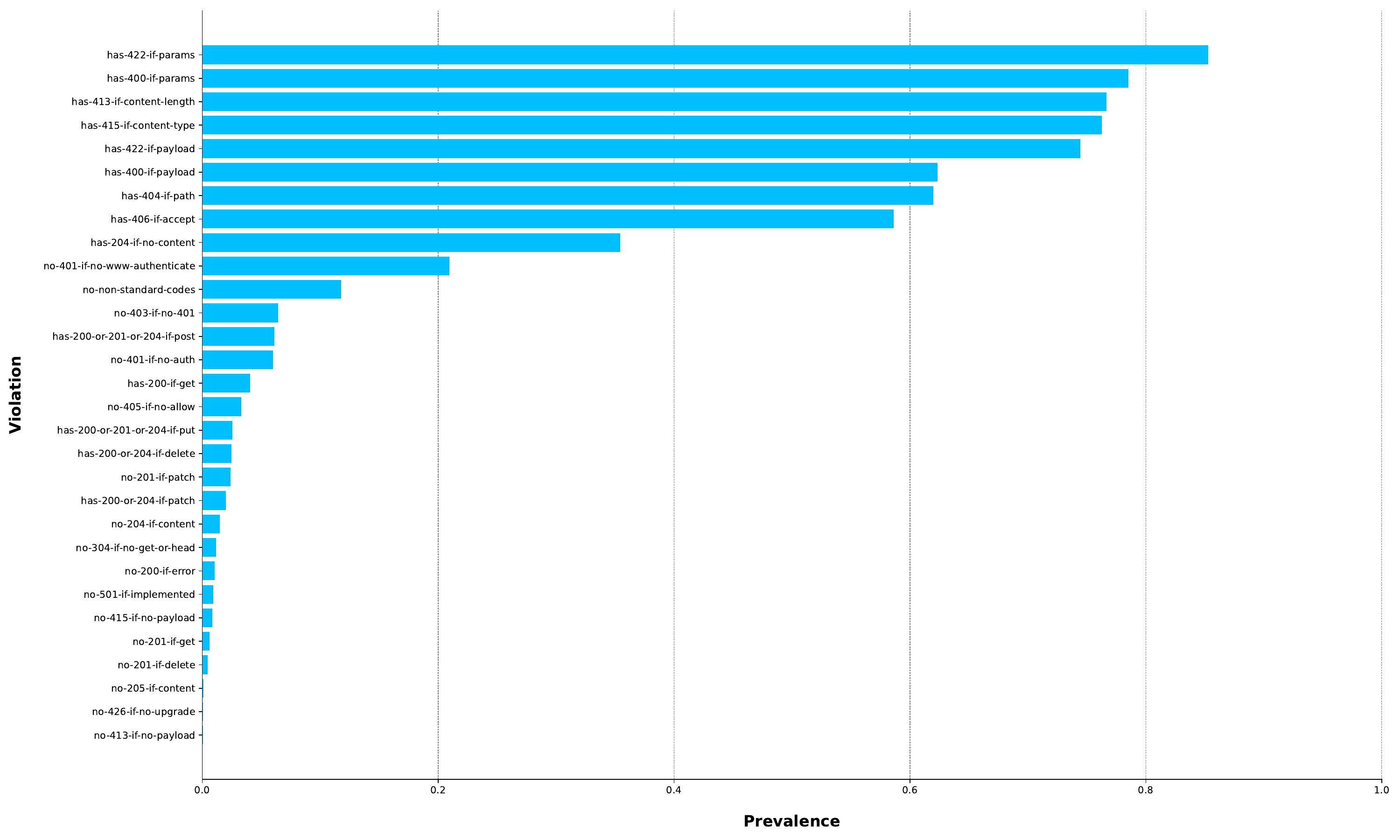}
    \caption{Distribution of status code misuse prevalence for the APIs.guru dataset.}
    \Description{Image of the distribution of status code misuse prevalence for the APIs.guru dataset.}
    \label{fig:chart-violation-prevalence-apisguru}
\end{figure}

\begin{figure}[t]
    \centering
    \includegraphics[width=0.9\linewidth]{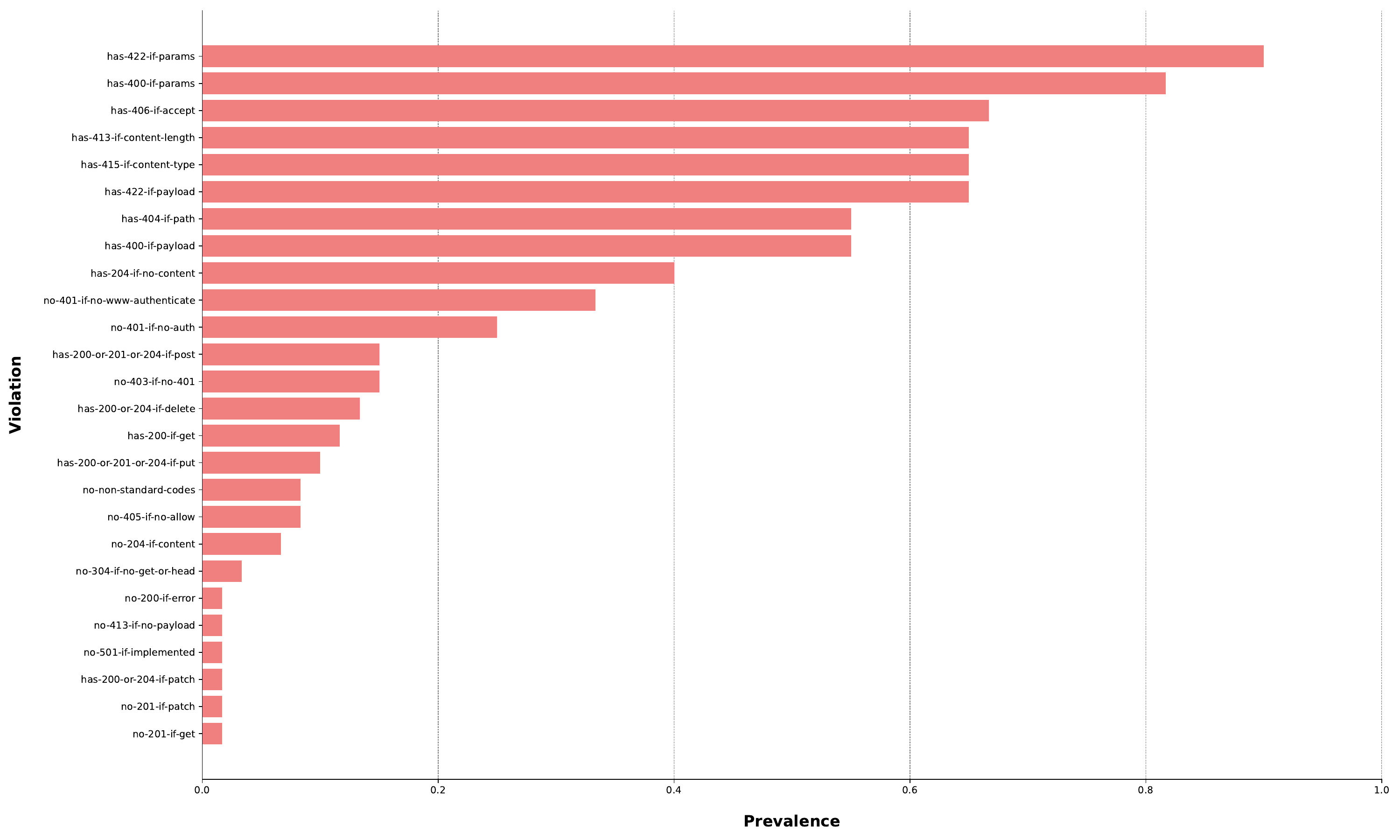}
    \caption{Distribution of status code misuse prevalence for the PRAB dataset.}
    \Description{Image of the distribution of status code misuse prevalence for the PRAB dataset.}
    \label{fig:chart-violation-prevalence-prab}
\end{figure}

\begin{figure}[t]
    \centering
    \includegraphics[width=0.9\linewidth]{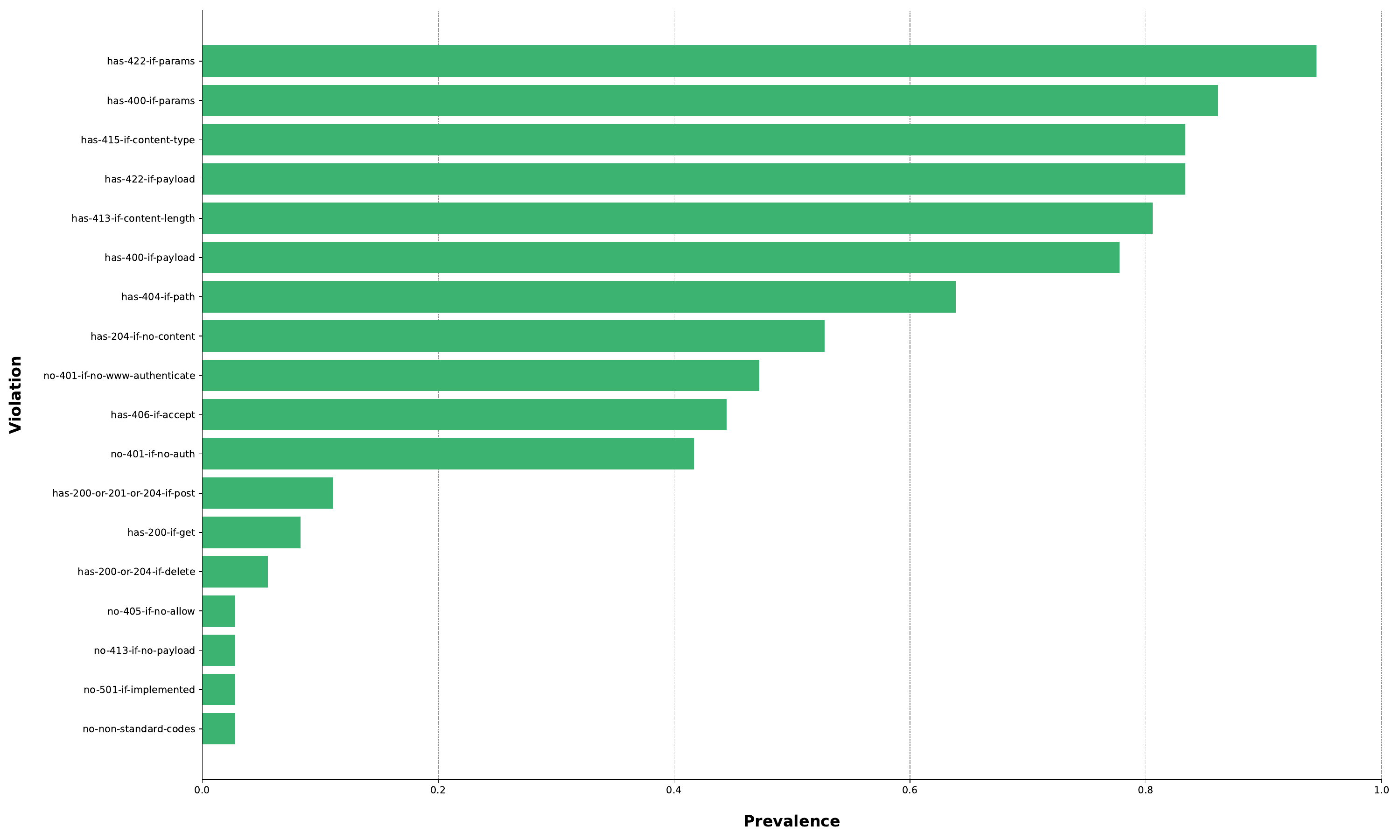}
    \caption{Distribution of status code misuse prevalence for the WFD dataset.}
    \Description{Image of the distribution of status code misuse prevalence for the WFD dataset.}
    \label{fig:chart-violation-prevalence-wfd}
\end{figure}

Doing so allows us to measure the violations which occur the most across APIs, independent of how many violations occur in a single API and thus avoiding size bias. We report the results for the three evaluation datasets. \Cref{fig:chart-violation-prevalence-apisguru} illustrates the prevalence for the APIs.guru dataset, \Cref{fig:chart-violation-prevalence-prab} illustrates the prevalence for the PRAB dataset, and \Cref{fig:chart-violation-prevalence-wfd} illustrates the prevalence for the WFD dataset. A total of 696,891 rule violations were found across 2,612 OpenAPI specifications, with \Cref{tab:static-results} presenting the results for each dataset.

\begin{table}[t]
    \centering
    \caption{Results of the static analysis for each dataset.}
    \label{tab:static-results}
    \small
    \begin{tabular}{l r r r r r}
        \toprule

        \textbf{Dataset} & \textbf{No. APIs} & \textbf{No. Rule Violations} & \textbf{Average per API} & \textbf{No. APIs Without Violations} \\

        \midrule

        APIs.guru & 2,516 & 669,485 & 266.09 & 2 \\

        PRAB & 60 & 19,837 & 330.62 & 0 \\

        WFD & 36 & 7,569 & 210.25 & 0 \\

        \bottomrule
    \end{tabular}
\end{table}

As shown, the most frequent rule violation across the three datasets is \texttt{has-422-if-params} (i.e., missing \texttt{422 Unprocessable Content} if there are parameters), suggesting a lack of implemented client error responses when there are query parameters. Across the datasets, we also observe that the most frequent rule violations are of the \texttt{has-4xx} category (e.g., \texttt{has-400-if-params}, \texttt{has-413-if-content-length}, \texttt{has-404-if-path}, etc.), suggesting a more broad lack of documented client errors for various contexts (e.g., missing \texttt{400 Bad Request} if there are query parameters, missing \texttt{413 Content Too Large} for the \texttt{Content-Length} header, missing \texttt{404 Not Found} if there is a path parameter, etc.). These overall results suggest that specific client errors are less documented.

We may also observe the following:

\begin{itemize}
    \item The most frequent rule violation for missing status codes from the \texttt{2xx} (success) range is \texttt{has-204-if-no-content} (across all datasets), suggesting that many APIs either forget to document response content or incorrectly use the \texttt{204 No Content} status code.

    \item Misuses related to security were found, especially \texttt{no-401-if-no-www-authenticate}. This suggests that developers often forget to document the \texttt{WWW-Authenticate} header when a \texttt{401 Unauthorized} status code response is used.

    \item More ``severe'' misuses were found, such as forgetting to document \texttt{200 OK} responses for \texttt{GET} methods (\texttt{has-200-if-get}), implementing \texttt{201 Created} responses for \texttt{PATCH} methods (\texttt{no-201-if-patch}, \texttt{PATCH} modifies data but does not create it), or using non-standard status codes (\texttt{no-non-standard-codes}). While such rule violations were less frequent across the datasets, this suggests that they are not exempt from REST APIs in practice.
\end{itemize}

Moreover, we observe that all API specifications from the PRAB and WFD datasets contained status code misuses. However, in the APIs.guru dataset, two API specifications (out of 2,516) did not trigger any status code usage rule violations, namely the \texttt{azure-com-dynamicstelemetry} and the \texttt{mastercard-com-bintableresource} API specifications. These results suggest that status code misuses are omnipresent and systematic (to a certain extent) in REST API specifications.

\begin{tcolorbox}
    \textbf{RQ.3 Summary}: After executing \statictool{} on 2,612 distinct OpenAPI specifications, we found a total of 696,891 status code misuses. Frequent misuses are related to undocumented client error (\texttt{4xx}) responses for specific cases (e.g., query/path parameters and headers). Security-related misuses were also found, along with misuses suggesting more ``severe'' misunderstandings of HTTP semantics (in fewer numbers). Misuses were found in 99.92\% of specifications (only two did not contain any misuses), revealing their omnipresence in REST API specifications.
\end{tcolorbox}

\subsection{RQ.4: Dynamic Analysis of Status Code Misuses in REST APIs}

As previously discussed in Section~\ref{subsec:dynamicanalysis}, it is not possible to directly translate all static status checks into a dynamic version. 
Of the 30 unique status code usage rules discussed in Section~\ref{subsec:rq2}, we could only translate 16 of them. 
Those are listed in Table~\ref{tab:dynamicrules}, including the type of inputs they require to be computed. 

\begin{table}[!t]
\centering
\caption{ 
\label{tab:dynamicrules}
List of implemented rules that can be checked dynamically, including the inputs they need to be computed.
}
\small
    \begin{tabular}{l l}
        \toprule
        \textbf{Identifier}  & \textbf{Inputs}\\
        \midrule
        \texttt{has-406-if-accept}  & \texttt{status}, \texttt{headers}, \texttt{schema} \\   
        \texttt{no-201-if-delete} & \texttt{status}, \texttt{verb} \\        
        \texttt{no-201-if-get} & \texttt{status}, \texttt{verb} \\    
        \texttt{no-201-if-patch} & \texttt{status}, \texttt{verb} \\
        \texttt{no-204-if-content} & \texttt{status}, \texttt{body} \\
        \texttt{no-205-if-content} & \texttt{status}, \texttt{body} \\
        \texttt{no-304-if-no-get-or-head} & \texttt{status}, \texttt{verb} \\        
        \texttt{no-401-if-no-auth} & \texttt{status}, \texttt{schema} \\
        \texttt{no-401-if-no-www-authenticate} & \texttt{status}, \texttt{headers} \\        
        \texttt{no-403-if-no-401}  & \texttt{status}, \texttt{schema} \\
        \texttt{no-405-if-no-allow} & \texttt{status}, \texttt{headers} \\        
        \texttt{no-413-if-no-payload} & \texttt{status}, \texttt{body} \\        
        \texttt{no-415-if-no-payload} & \texttt{status}, \texttt{body} \\
        \texttt{no-426-if-no-upgrade} & \texttt{status}, \texttt{headers} \\
        \texttt{no-501-if-implemented} & \texttt{status} \\        
        \texttt{no-non-standard-codes} & \texttt{status} \\
        \bottomrule
    \end{tabular}
\end{table}

At a high level, all the \texttt{no-*} rules could be adapted, and no \emph{has-*}, with two exceptions.
First, in the case of \texttt{has-406-if-accept}, we can check that, if a fuzzer is sending requests valid according to the schema (e.g., no robustness testing of invalid inputs), then it should never happen that the API responds with a \texttt{406} status code.
Note that, this is not the same as checking that \texttt{406} is declared in the schema for endpoints that have input payloads. 
We use the same label \texttt{has-406-if-accept} for simplicity and consistency with the static rules, but a different name could have been coined for this rule specific to the dynamic case.
Second, we did not define a dynamic version of \texttt{no-200-if-error}, as it is not really viable to reliably compute in a dynamic context.
Determining whether a response represents an error message is not a simple, deterministic task. 
Natural language processing (NLP), and techniques like LLM, could be used for fixed, static files like an OpenAPI schema.
However, in a dynamic fuzzing session where hundreds of thousands (if not millions) of HTTP calls are made, applying NLP/LLM might not be viable to analyze each single obtained HTTP response.

We implemented all these oracles in \evo. 
To first verify their implementation correctness, we created an artificial API, using Kotlin and SpringBoot, with one faulty endpoint for each of these 16 rules.
This API is then used inside \evo as an E2E test for \evo itself, which is executed in CI at each new code change~\cite{arcuri2023building}.

All these injected faults can be reliably detected, but we could not produce faults for two rules, specifically \texttt{no-204-if-content} and \texttt{no-205-if-content}. 
Trying to return an HTTP response with a body payload and a status code of either \texttt{204} or \texttt{205} is not possible in SpringBoot (with its default Tomcat as HTTP server). 
If a developer tries to do that (as we did in our artificial API), the framework silently remove the payload from the response (with no error message or thrown exception).
Some HTTP frameworks might prevent HTTP mistakes (e.g., presence of body payloads in \texttt{204}/\texttt{205} responses). 
But, the choice of which HTTP rules to enforce or not in these frameworks seem rather arbitrary.  

Another interesting case is \texttt{no-non-standard-codes}.
Internally, \evo uses Jersey\footnote{https://github.com/eclipse-ee4j/jersey} 
to make HTTP requests. 
If a response with status code below \texttt{100} is returned, Jersey crashes and throw an exception.
However, it creates an HTTP response object, without throwing any exception, if the status code is equal or above \texttt{600}. 
The problem is then in the generated tests, which can use different libraries to make HTTP calls (especially when outputting test suite in different programming languages such as JavaScript and Python, both supported by \evo~\cite{zhang2023javascript,arcuri2025widening}). 
For example, in the case of RestAssured, in contrast to Jersey it crashes in both cases, i.e., also for values equal or above \texttt{600}. 
Note: to handle these issues, in \evo those HTTP calls returning a status code out of range are simply put inside a try/catch block. 
Still, it is important to highlight how the specifications of HTTP can be interpreted differently between different libraries and frameworks.

Once these 16 rules were implemented in \evo, we carried out an experiment on the 36 APIs currently in WFD. 
On each API, we ran \evo for 1 hour, using its white-box mode. 
To take into account the randomness of the employed algorithms, each experiment was repeated 10 times, for a total of 360 distinct runs. 
As we are just interested to see which of our novel rules could detect faults, for reason of space we do not report any statistics on achieved code coverage or schema coverage. 
Table~\ref{tab:results-wb} shows the results of these experiments. 
The number of faults is counted based on the number of distinct endpoints (path+verb) that trigger our rules. 

\begin{table}[t]
\centering
\caption{ 
Detected faults on the 36 APIs from WFD:
F952 (\texttt{no-201-if-get}),
F955 (\texttt{no-413-if-no-payload}),
F957 (\texttt{no-401-if-no-auth}),
F958 (\texttt{no-403-if-no-401}),
F959 (\texttt{has-406-if-accept}),
F961 (\texttt{no-401-if-no-authenticate}),
F962 (\texttt{no-405-if-no-allow}),
and
F963 (\texttt{no-501-if-implemented}).
\label{tab:results-wb}
}
\small
\begin{tabular}{lrrrrrrrrr}
\toprule 
 \textbf{SUT} & \textbf{\#Endp} & \textbf{F952} & \textbf{F955} & \textbf{F957} & \textbf{F958} & \textbf{F959} & \textbf{F961} & \textbf{F962} & \textbf{F963}  \\ 
\midrule 
\emph{bibliothek} & 8 &  &  &  &  &  &  &  &  \\ 
\rowcolor{gray!30} \emph{blogapi} & 52 &  &  & 32.4 & 3.1 &  & 32.4 &  &  \\ 
\emph{catwatch} & 14 &  &  & 2.0 &  &  &  &  &  \\ 
\rowcolor{gray!30} \emph{cwa-verification} & 5 &  &  &  &  &  &  &  &  \\ 
\emph{erc20-rest-service} & 13 &  &  &  &  &  &  &  &  \\ 
\rowcolor{gray!30} \emph{familie-ba-sak} & 183 &  &  &  & 171.0 &  & 162.0 &  &  \\ 
\emph{features-service} & 18 &  &  &  &  &  &  &  &  \\ 
\rowcolor{gray!30} \emph{genome-nexus} & 23 &  &  &  &  &  &  &  &  \\ 
\emph{gestaohospital} & 20 &  &  &  &  &  &  &  &  \\ 
\rowcolor{gray!30} \emph{http-patch-spring} & 6 &  &  &  &  &  &  &  &  \\ 
\emph{languagetool} & 2 &  &  &  &  &  &  &  &  \\ 
\rowcolor{gray!30} \emph{market} & 13 &  &  & 7.6 &  & 3.9 & 6.6 &  &  \\ 
\emph{microcks} & 88 &  &  & 32.0 & 5.0 &  &  &  &  \\ 
\rowcolor{gray!30} \emph{ocvn} & 258 &  &  &  &  & 0.9 &  &  &  \\ 
\emph{ohsome-api} & 134 &  & 0.2 &  &  &  &  &  &  \\ 
\rowcolor{gray!30} \emph{pay-publicapi} & 10 &  &  &  &  &  & 10.0 &  &  \\ 
\emph{person-controller} & 12 &  &  &  &  &  &  &  &  \\ 
\rowcolor{gray!30} \emph{proxyprint} & 74 &  &  & 52.3 &  &  & 3.3 &  &  \\ 
\emph{quartz-manager} & 11 &  &  &  &  &  & 11.0 &  &  \\ 
\rowcolor{gray!30} \emph{reservations-api} & 7 &  &  &  &  &  & 6.0 &  &  \\ 
\emph{rest-ncs} & 6 &  &  &  &  &  &  &  &  \\ 
\rowcolor{gray!30} \emph{rest-news} & 7 &  &  &  &  &  &  &  &  \\ 
\emph{rest-scs} & 11 &  &  &  &  &  &  &  &  \\ 
\rowcolor{gray!30} \emph{restcountries} & 22 &  &  &  &  &  &  &  &  \\ 
\emph{scout-api} & 49 &  &  & 34.0 & 21.2 &  &  &  &  \\ 
\rowcolor{gray!30} \emph{session-service} & 8 &  &  &  &  &  &  &  &  \\ 
\emph{spring-actuator-demo} & 2 &  &  &  &  &  &  &  &  \\ 
\rowcolor{gray!30} \emph{spring-batch-rest} & 5 &  &  &  &  &  &  &  &  \\ 
\emph{spring-ecommerce} & 27 &  &  & 18.8 &  & 0.2 & 18.8 &  &  \\ 
\rowcolor{gray!30} \emph{spring-rest-example} & 9 &  &  &  &  &  &  &  & 1.0 \\ 
\emph{swagger-petstore} & 19 &  &  &  &  &  &  & 0.6 &  \\ 
\rowcolor{gray!30} \emph{tiltaksgjennomforing} & 79 &  &  &  & 18.6 &  & 72.0 &  &  \\ 
\emph{tracking-system} & 67 &  &  &  &  &  &  &  &  \\ 
\rowcolor{gray!30} \emph{user-management} & 21 & 1.0 &  &  &  &  &  &  &  \\ 
\emph{webgoat} & 204 &  &  & 0.9 &  &  & 0.9 &  &  \\ 
\rowcolor{gray!30} \emph{youtube-mock} & 1 &  &  &  &  &  &  &  &  \\ 
\midrule 
\textbf{Mean}  & 41 & 0.0 & 0.0 & 5.0 & 6.1 & 0.1 & 9.0 & 0.0 & 0.0 \\ 
\textbf{Median}  & 14 & 0.0 & 0.0 & 0.0 & 0.0 & 0.0 & 0.0 & 0.0 & 0.0 \\ 
\textbf{Sum}  & 1488 & 1.0 & 0.2 & 179.9 & 218.9 & 5.0 & 322.9 & 0.6 & 1.0 \\ 
\bottomrule 
\end{tabular} 

\end{table}

Note that we used WFD as it is. 
We did not inject any manual fault in any of these APIs. 
What found are actual real faults in these APIs, unknown to us before running these experiments. 

Based on the results shown in Table~\ref{tab:results-wb}, it was possible to detect 8 out 16 types of faults in these APIs. 
To distinguish them, and enable comparisons between different fuzzers in the future,
these faults are currently marked with Web Fuzzing Commons (WFC)~\cite{sahin2025wfc} codes (where the \texttt{9XX} range is for temporary codes not part of the proposed standard yet). 

One of the main advantages of state-of-the-art fuzzers such as \evo, it is that it can generate executable test cases that can be used to help investigating and debugging any found fault. 
It is not our goal here to go into the details of each of the hundreds of different faults reported in Table~\ref{tab:results-wb}. 
But let us discuss in more details some of the examples we found most interesting.

\begin{figure}
\begin{lstlisting}[language=java]
/**
* Calls:
* (201) GET:/users/rbac/salt
* Found 2 potential faults. Type-codes: 101, 952
*/
@Test(timeout = 60000)
public void test_45_getOnSaltShowsFaults_101_952() throws Exception {
        
    // Fault101. Received A Response From API With A Structure/Data That Is Not Matching Its Schema. Type: validation.response.status.unknown Response status 201 not defined for path '/users/rbac/salt'.
    // Fault952. HTTP/REST-Design Violation: no-201-if-get.
    given().accept("*/*")
        .header("x-EMextraHeader123", "")
        .get(baseUrlOfSut + "/users/rbac/salt?EMextraParam123=42")
        .then()
        .statusCode(201)
        .assertThat()
        .contentType("text/plain")
        .body(containsString("zmIxrvlmrCMbVOatDAcaySfFtHRRrW7f"));
}
\end{lstlisting}
\caption{\label{fig:evo201}
Generated test for \emph{user-management} API showing a F952 \texttt{no-201-if-get} detected fault.
}
\Description{Figure}
\end{figure}

Figure~\ref{fig:evo201} shows an example of test generated by \evo on the \emph{user-management} API. 
Here, a \texttt{GET} requests is wrongly marked as creating a resource (i.e., status code \texttt{201}). 
Based on  what  the semantics of that endpoint would seem, a \texttt{POST} should had rather being used. 
This is not just a matter of ``design'', but also it has functional consequences, as \texttt{GET} is idempotent (i.e., any idempotent request could be repeated one or several times by any link between the client and the server without any feedback to the client).

\begin{figure}
\begin{lstlisting}[language=java]
/**
* Calls:
* (413) POST:/elements/count
* Found 2 potential faults. Type-codes: 101, 955
*/
@Test @Timeout(60)
public void test_14_postOnCountShowsFaults_101_955() throws Exception {
        
    // Fault101. Received A Response From API With A Structure/Data That Is Not Matching Its Schema. Type: validation.response.body.unexpected No response body is expected but one was found.
    // Fault955. HTTP/REST-Design Violation: no-413-if-no-payload.
    given().accept("application/json")
        .header("x-EMextraHeader123", "")
        .header("X-REQUEST-URI", "_EM_54169_XYZ_")
        .noContentType()
        .post(baseUrlOfSut + "/elements/count?" + 
                "bboxes=8.67%2C49.39%2C8.71%2C49.42&" + 
                "format=json&" + 
                "timeout=-7.700129630469112E8&" + 
                "keys=2RIKSBBuOf8FHAH&" + 
                "values=_EM_17167_XYZ_")
        .then()
        .statusCode(413)
        .assertThat()
        .contentType("application/json")
        .body("'status'", numberMatches(413))
        .body("'message'", containsString("The given query is too large in respect to the given timeout. Please use a smaller region and/or coarser time period."))
        .body("'requestUrl'", containsString("_EM_54169_XYZ_?bboxes=8.67%2C49.39%2C8.71%2C49.42&format=json&timeout=-7.700129630469112E8&keys=2RIKSBBuOf8FHAH&values=_EM_17167_XYZ_"));
    }
\end{lstlisting}
\caption{\label{fig:evo413}
Generated test for \emph{ohsome-api} showing a F955 \texttt{no-413-if-no-payload} detected fault.
}
\Description{Figure}
\end{figure}

Figure~\ref{fig:evo413} shows a test generated for \emph{ohsome-api} in which a status code \texttt{413 Content Too Large} is returned for a request that has no body. 
The status code \texttt{413} is meant for request \emph{contents} that are too large, and not for representing internal computations that might be too time consuming based on the given input parameters. 
Note that the term \emph{content} has a well defined, formal meaning in HTTP (and that is also the reason why the specifier ``Payload Too Large'' in RFC 7231 was changed into the more precise, less ambiguous ``Content Too Large'' in RFC 9110).
For this endpoint, a different status code than \texttt{413} should had been used. 

\begin{figure}
\begin{lstlisting}[language=java]
/**
* Calls:
* (501) GET:/api/problem/{code}
* Found 2 potential faults. Type-codes: 101, 963
*/
@Test @Timeout(60)
public void test_5_getOnProblemShowsFaults_101_963() throws Exception {
        
    // Fault101. Received A Response From API With A Structure/Data That Is Not Matching Its Schema. Type: validation.response.status.unknown Response status 501 not defined for path '/api/problem/{code}'.
    // Fault963. HTTP/REST-Design Violation: no-501-if-implemented.
    given().accept("application/json")
            .header("x-EMextraHeader123", "")
            .get(baseUrlOfSut + "/api/problem/_EM_13918_XYZ_")
            .then()
            .statusCode(501)
            .assertThat()
            .contentType("application/json")
            .body("'message'", containsString("No data with this code"))
            .body("'details'", containsString("uri=/api/problem/_EM_13918_XYZ_"));
}
\end{lstlisting}
\caption{\label{fig:evo501}
Generated test for \emph{spring-rest-example} API showing a F963 \texttt{no-501-if-implemented} detected fault.
}
\Description{Figure}
\end{figure}

A similar issue can be seen in Figure~\ref{fig:evo501}. 
Here, a wrong error status code is returned on a request based on its input (the path element \texttt{\{code\}} in this case),  where a \texttt{501} is incorrect to return for an endpoint that is defined and implemented in the API (\texttt{no-501-if-implemented}).
Based on the returned error message, likely here returning a 404 would had been more appropriate.

\begin{figure}
\begin{lstlisting}[language=java]
/**
    * Calls:
    * (401) DELETE:/api/vilkaarsvurdering/{behandlingId}/{vilkaarId}
    * Found 1 potential fault of type-code 961
    */
    @Test @Timeout(60)
    public void test_138_deleteOnVilkaarsvurd401MissingWwwAuthenticate() throws Exception {
        
        // Fault961. HTTP/REST-Design Violation: no-401-if-no-authenticate.
        given().accept("*/*")
                .header("x-EMextraHeader123", "")
                .contentType("application/json")
                .body(" \"_EM_14_XYZ_\" ")
                .delete(baseUrlOfSut + "/api/vilkaarsvurdering/428/8697413960304443323?EMextraParam123=_EM_15_XYZ_")
                .then()
                .statusCode(401)
                .assertThat()
                .header("www-authenticate", isEmptyOrNullString())
                .contentType("application/json")
                .body("'data'", nullValue())
                .body("'status'", containsString("FEILET"))
                .body("'melding'", containsString("Unauthorized"))
                .body("'frontendFeilmelding'", nullValue())
                .body("'stacktrace'", nullValue());
    }
\end{lstlisting}
\caption{\label{fig:evo401}
Generated test for \emph{familie-ba-sak} API showing a F961 \texttt{no-401-if-no-authenticate} detected fault.
}
\Description{Figure}
\end{figure}

Finally, let us consider the case of failed authentication in Figure~\ref{fig:evo401}, from the API \emph{familie-ba-sak} from the Norwegian Labour and Welfare Administration (NAV). 
Here, on a \texttt{401} non-authenticated response, the API does not satisfy the HTTP protocol, as it does not provide any entry for the \texttt{WWW-Authenticate} HTTP header on \texttt{401} responses (\texttt{no-401-if-no-authenticate}). 
In HTTP, this is not optional, it is mandatory. 
Note that here, to make the test cases more readable and more usable for the final test engineers, we add checks on returned HTTP headers, but only for the cases that need it, like \texttt{WWW-Authenticate} for \texttt{no-401-if-no-authenticate}, 
\texttt{Allow} for \texttt{no-405-if-no-allow}
and 
\texttt{Upgrade} for \texttt{no-426-if-no-upgrade}.

\begin{tcolorbox}
    \textbf{RQ.4 Summary}: On the 36 APIs of WFD, our novel oracles were able to detect hundreds of faulty endpoints, finding 8 different types of HTTP faults. 
\end{tcolorbox}
\section{Discussion}
\label{sec:discussion}

\subsection{Static and Dynamic Analysis Value}

Our results highlight that static and dynamic approaches may complement one another for detecting status code misuses in REST APIs. As static analysis operates on the API specification level (i.e., analysis of OpenAPI specifications), it can identify inconsistencies between the documented behavior and established status code usage guidelines without requiring access to a running implementation. This makes it useful for detecting misuses at the specification level, such as inappropriate or missing status codes, and for identifying issues early in the API development lifecycle. However, static analysis is limited to the behaviors declared in the specification and cannot determine if documented responses are actually implemented by the API server.

In contrast, dynamic analysis observes the behavior of API implementations. Consequently, it can detect status code misuses that are not represented in API specifications, including responses that are returned by the implementation but are not declared in the corresponding schema. For example, \Cref{fig:evo201} shows a misuse detected through dynamic analysis that was not identified by static analysis. The explanation is that the corresponding status code is returned by the implementation but is not declared in the API specification (i.e., only \texttt{200}, \texttt{401}, \texttt{403}, \texttt{404} are declared in the specification, but not \texttt{201}). More generally, this illustrates that API implementations may exhibit behaviors that are not captured by their corresponding specifications. Similarly, a status code declared in an API specification does not necessarily imply that the server actually uses or implements that response.

Overall, these observations suggest that static and dynamic analysis should complement one another rather than compete:

\begin{itemize}
    \item Static analysis provides coverage of API documentation, and may identify potential status code misuses without execution of the system.
    \item Dynamic analysis provides evidence about the behavior of the deployed implementation, and may uncover undocumented and/or specification-inconsistent behaviors.
\end{itemize}

Combining both approaches provides a more complete look at status code usage and misuses across both the intended API specification (i.e., contract) and its actual implementation.

\subsection{Implications for API Testers}

REST API testing tools rely on status code ranges for test outcomes and to find server errors: \texttt{2xx} for valid requests, \texttt{4xx} for client errors, and \texttt{5xx} for server errors. Golmohammadi, Zhang, and Arcuri~\cite{golmohammadi2023testing} highlight that \texttt{5xx} status codes are used to identify faults in over 30 different works on REST API testing. Thus, misusing status codes can lead to false positives/negatives in such tools. For instance, if a server responds with \texttt{5xx} status codes for client errors, this leads to false positives (inaccurate bug report). This misleads the tester, as this false server error is unlikely to exhibit an interesting behavior. Similarly, if an API uses \texttt{2xx} or \texttt{4xx} status codes to represent server errors, this leads to false negatives.

This can also happen with client errors if the API always responds with \texttt{2xx} status codes. For instance, when sending a request with an invalid path to the Deezer API, a \texttt{404 Not Found} client error is expected. However, the API replies with \texttt{200 OK}. While this status code indicates a success, an error message is found in the response body with an API-defined \texttt{600} code. This deviates from REST API design (non-standard status codes), and confuses tools if no additional response parsing is performed. Similarly, false positives can be observed in popular API frameworks such as Spring. For instance, the Spring Petclinic REST API returns \texttt{500 Internal Server Error} status codes when a path does not exist (i.e., \texttt{NoResourceFoundException}), instead of \texttt{404 Not Found}. While it is possible to specify status code mappings in Spring, they can be omitted and thus the \texttt{500} code is used as a fallback.

To illustrate the prevalence of status code reliance in API testing tools, we analyzed the oracles of state-of-the-art testing tools for REST APIs. We selected open-source testing tools appearing in a survey on REST API testing~\cite{golmohammadi2023testing}, along with other recent tools \cite{karlsson2020quickrest,liu2022icse}. Table \ref{tab:testing-tools} presents the testing tools and their respective oracles. As shown, some tools incorporate additional checks in their oracles, such as specification conformance, response validation, checkers, etc. However, all tools do rely on status codes (i.e., \texttt{5xx} server error codes) to find bugs. While this is perfectly acceptable as \texttt{5xx} codes represent server errors according to official HTTP standards, it highlights one possible detrimental effect of misusing status codes.

\begin{table}[t]
    \centering
    \caption{State-of-the-art testing tools for REST APIs and their respective oracles.}
    \label{tab:testing-tools}
    \small
    \begin{tabular}{l  l  l}
        \toprule
    
        \textbf{Tool Name} & \textbf{Reference} & \textbf{Oracle}\\
    
        \midrule

        bBOXRT & \cite{laranjeiro2021black} & Status codes, responses, test metadata \\

        QuickREST & \cite{karlsson2020quickrest} & Status codes, specification conformance \\

        Morest & \cite{liu2022icse} & Status codes \\

        Restats & \cite{corradini2021restats} & Status codes \\

        RestCT & \cite{wu2022combinatorial} & Status codes \\
    
        RESTler & \cite{atlidakis2019restler} & Status codes, predefined checkers \\

        RESTest & \cite{martin2021restest} & Status codes, response validation \\

        RestTestGen & \cite{viglianisi2020resttestgen} & Status codes, schema match \\

        Schemathesis & \cite{hatfield2022deriving} & Status codes, semantic checks \\
    
        \bottomrule
    \end{tabular}
\end{table}

\begin{tcolorbox}
\textbf{Recommendations for Testers}: REST API testing tools should analyze responses in depth to avoid false positives/negatives from status codes. Response messages should be parsed to check for errors, and status code mappings should be set up when APIs implement their own codes.
\end{tcolorbox}

\subsection{Implications for API Users}

Another implication of misusing status codes is that API clients are prone to receiving ambiguous responses from servers. For instance, in a microservice architecture with loosely coupled REST APIs, the services would rely on HTTP standards to understand responses. However, if a service always responds with \texttt{200 OK} status codes, other services following HTTP standards would interpret all requests as valid. Doing so could cause various problems, such as rendering errors on pages, propagating invalid responses to downstream services, or logging misleading success messages.

Status code misuses also introduce other client-related problems. Without precise client errors codes from the server, debugging invalid requests becomes a difficult task. For instance, a server could respond with overly generic \texttt{400 Bad Request} status codes, instead of more specific codes such as \texttt{401 Unauthorized} (for a lack of authentication) or \texttt{413 Content Too Large} (for unsupported content size). Moreover, front-end applications often use status codes to display messages (e.g., “Saved successfully” for \texttt{201 Created}). If the wrong code is used, users see misleading feedback.

In consequence, handling status code misuses is also relevant for REST API clients (e.g., users and dependent microservices).

\begin{tcolorbox}
\textbf{Recommendations for Clients}: REST API clients should not directly trust status codes. Instead, clients should check documentation for potential API-dependent codes or API-defined behaviors. Similarly to testers, clients should also analyze response content and messages in-depth.
\end{tcolorbox}

\section{Threats to Validity}
\label{sec:threats}

\subsection{Internal Validity}

\paragraph{Status Code Usage Rules}
The ground truth for status code usage rules was derived through manual analysis, which may be prone to human errors or subjective disagreements about what constitutes a status code misuse. To mitigate this threat, we analyzed the HTTP standards (status codes and methods) defined in the official IANA registry and collected the distribution of status codes across individual REST APIs.

\paragraph{Tool Implementation}
The implementation of our tools (i.e., \statictool{} and \evo) may be prone to issues or bugs. We mitigated this threat with validation, testing, and by cross-checking outputs.

\subsection{External Validity}

\paragraph{Dataset Representativeness}
The benchmark of REST API specifications may not fully reflect the diversity of real-world API specifications, which introduces potential construction bias. To mitigate this threat, we utilized three different datasets (cf. \Cref{subsubsec:dataset}) with APIs varying in popularity, size, and application domain. Moreover, it is probable that some APIs from the datasets may have been updated since our evaluation, creating inconsistencies between the specification versions analyzed and their current state.

\paragraph{Tool Practicality}
To mitigate threats related to the practical usefulness of our tool, we designed \statictool{} to be lightweight, being able to run on a ``standard'' laptop in a couple of seconds. Moreover, we opted for input/outputs in JSON, HTML, and LOG file formats, which are widely used in practice.

\paragraph{Specification Formats}
Our study was limited to using the OpenAPI Specification format to document REST APIs (for the static analysis, cf. \Cref{subsec:scoas}). Although this format is the most widely used, we acknowledge this as a limitation regarding other existing documentation formats for REST APIs (e.g., Postman Collections, RAML, API Blueprint).

\subsection{Construct Validity}

\paragraph{Misuse Definition}
A potential threat to construct validity is the definition of a ``status code misuse''. As a result, the notion of misuse used in this work may not fully capture all context-dependent or domain-specific interpretations, potentially introducing subjectivity into the evaluation.
\section{Conclusion and Future Work}
\label{sec:conclusion}

In this paper, we explored the static and dynamic detection of HTTP status code misuses in REST APIs. To do so, we implemented a set of \nbRules{} status code usage rules based on an exhaustive analysis of HTTP semantics and REST API principles. We then operationalized the rules by implementing a static analyzer (i.e., \statictool{}) and a dynamic analyzer (i.e., an extension of \evo). We evaluated the effectiveness of the two tools with an evaluation dataset combining three state-of-the-art REST API benchmarks: APIs.guru, PRAB, and WFD. Our evaluation demonstrated that status code misuses are frequent and systematic in REST APIs, and that the static and dynamic approaches should be used in a complementary manner. We also provided insight for API users and testers, suggesting that REST API responses should always be analyzed in-depth and that a sole reliance on HTTP status codes in insufficient.

For future work, we plan to identify which status code usage rules are more severe based on their misuse implications in practice. We also plan to improve the tools by automatically fixing the detected status code misuses. However, this is challenging for missing status codes, as their response implementation depends on a specific context only known by the developer (e.g., if a \texttt{2xx} status code is missing in a \texttt{GET} request, either \texttt{200 OK}, \texttt{204 No Content}, or both may be implemented). However, for REST APIs comprising of a specification and an implementation, this might be feasible as context may be extracted from the implementation. We would also like to assess the security implications of status code misuses (e.g., in microservice systems).

\section*{Data Availability}
Both SCOAS\footnote{\url{https://github.com/alixdecr/scoas}} and \evo\footnote{\url{https://github.com/webfuzzing/evomaster}} (along with their evaluation data) are publicly available and open-source on GitHub.
Each new release of \evo is also automatically uploaded to Zenodo, such as the v6.1.1~\cite{zenodo611evomaster} used in this study.

\section*{Disclosure}
The authors have no competing interests to declare that are relevant in the context of this paper. Generative AI was used for limited rephrasing purposes. The authors reviewed the final content and take full responsibility for it.

\begin{acks}
Gilles Perrouin is an FNRS Research Associate. This research was partially funded by the CyberExcellence by DigitalWallonia project (No. 2110186), funded by the Public Service of Wallonia (SPW Recherche). 
Andrea Arcuri has been funded by the European Research Council (ERC) under the European Union’s Horizon 2020 research and innovation programme (EAST project, grant agreement No. 864972).
\end{acks}

\bibliographystyle{ACM-Reference-Format}
\bibliography{library/references}

\end{document}